\documentclass[12pt]{article}
\usepackage[a4paper,textwidth=14cm,textheight=24cm]{geometry}
\usepackage{amsmath,amsthm,amssymb,stmaryrd,wasysym}
\usepackage{color}
\usepackage{ulem}
\usepackage{stackrel}
\usepackage{graphicx}
\usepackage[all]{xy}

\input amssym.def

\def\be{\begin{equation}}
\def\ee{\end{equation}}
\def\bqn{\begin{eqnarray}}
\def\eqn{\end{eqnarray}}

\newcommand\bprop[1]{\begin{prop}#1\end{prop}}

\newcommand\bthm[2]{\begin{thm}[#1]#2\end{thm}}

\newcommand{\bea}{\begin{eqnarray}}
\newcommand{\eea}{\end{eqnarray}}
\newcommand{\p}{\partial}

\newcommand{\un}{^{-1}}
\newcommand{\bi}{\begin{itemize}}
\newcommand{\ei}{\end{itemize}}

\newcommand{\nn}{\nonumber}
\newcommand{\pl}{\left(}
\newcommand{\pr}{\right)}
\newcommand{\crl}{\left[}
\newcommand{\crr}{\right]}
\newcommand{\tl}{\left|}
\newcommand{\tr}{\right|}

\newcommand{\g}{\bar{g}}

\newcommand{\na}{\bar{\nabla}}
\newcommand{\R}{\bar{R}}

\newcommand{\half}{\frac{1}{2}}

\newcommand{\Lag}{\mathcal{L}}

\newcommand{\bU}{\bar{U}}

\newcommand{\bA}{\bar{A}}
\newcommand{\bg}{\bar{g}}

\newcommand{\dx}{\dot{x}}

\newcommand{\ddx}{\ddot{x}}
\newcommand{\ddu}{\ddot{u}}

\newcommand{\du}{\dot{u}}

\newcommand{\dep}{\pl t,x^i\pr}
\newcommand{\Om}{\Omega}
\newcommand{\om}{\omega}
\newcommand{\Omu}{\Omega^{-1}}

\newcommand{\dxita}{\frac{d x^i}{d\tau}}
\newcommand{\dxjta}{\frac{d x^j}{d\tau}}

\newcommand{\dxata}{\frac{d x^\alpha}{d\tau}}
\newcommand{\dxbta}{\frac{d x^\beta}{d\tau}}
\newcommand\der[1]{\frac{d #1}{d\tau}}

\newcommand{\ie}{\textit{i.e.}\, }
\newcommand{\eg}{\textit{e.g.}\, }

\theoremstyle{definition}

\newlength{\blength}
\renewcommand{\proof}[1]{\vspace{-.05cm}
\begin{list}{\bf Proof:}
{\listparindent=\parindent\parsep=0pt \labelwidth=-0.5cm
\labelsep=\parindent \addtolength{\labelsep}{-\blength}
\addtolength{\labelsep}{1.5cm}
\itemindent=-\blength
\addtolength{\itemindent}{\parindent} \leftmargin=1.0cm}
\item
#1~$\qedsymbol$\end{list}
\vspace{.0cm}}

\begin{document}

\thispagestyle{empty}


 \begin{centering}

{\large {\bfseries Embedding nonrelativistic physics\\\vspace{1mm}
inside a gravitational wave}\\\vspace{2mm}
}

 \vspace{2cm}
Xavier Bekaert \& Kevin Morand\\
\vspace{2mm}
{\small Laboratoire de Math\'ematiques et Physique Th\'eorique}\\
{\small Unit\'e Mixte de Recherche $7350$ du CNRS}\\
{\small F\'ed\'eration de Recherche $2964$ Denis Poisson}\\
{\small Universit\'e Fran\c{c}ois Rabelais, Parc de Grandmont}\\
{\small 37200 Tours, France} \\
\vspace{1mm}{\tt \footnotesize Xavier.Bekaert@lmpt.univ-tours.fr}\\
\vspace{1mm}{\tt \footnotesize Kevin.Morand@lmpt.univ-tours.fr}

\vspace{1.2cm}

\end{centering}

\begin{abstract}
Gravitational waves with parallel rays are known to have remarkable properties:
their orbit space of null rays possesses the structure of a nonrelativistic spacetime of codimension-one.
Their geodesics are in one-to-one correspondence with dynamical trajectories of a nonrelativistic system.
Similarly, the null dimensional reduction of Klein-Gordon's equation on this class of gravitational waves leads to a Schr\"odinger equation on curved space.
These properties are generalized to the class of gravitational waves with a null Killing vector field, of which
we propose a new geometric definition, as conformally equivalent to the previous class and such that the Killing vector field
is preserved.
This definition is instrumental for performing this generalization, as well as various applications. In particular, results on geodesic completeness are extended in a similar way.
Moreover, the classification of the subclass with constant scalar invariants is investigated.
\end{abstract}

\vspace{1.5cm}


\vspace{.5cm}


\vspace{4.5cm}

\pagebreak

\tableofcontents

\pagenumbering{arabic}

\pagebreak

\theoremstyle{plain}
\newtheorem{thm}{Theorem}[section]
\newtheorem{lem}[thm]{Lemma}
\newtheorem{cons}[thm]{Consequence}
\newtheorem{cor}[thm]{Corollary}
\newtheorem{prop}[thm]{Proposition}
\newtheorem{defprop}[thm]{Definition-Proposition}
\newtheorem{conj}[thm]{Conjecture}
\newtheorem{cl}[thm]{Claim}
\newtheorem{defi}[thm]{Definition}
\theoremstyle{definition}

\theoremstyle{remark}
\newtheorem{exa}[thm]{Example}
\newtheorem{rem}[thm]{Remark}

\section{Introduction}

With the advent of general relativity, the profound interaction between the geometry of spacetime and the motion of matter became a truism of modern physics, celebrated in the famous quote\footnote{``Space tells matter how to move. Matter tells space how to curve.''\cite{MTW}} of Wheeler. However, the intimate relationship between spacetime geometry and matter motion is by no means restricted to relativistic physics.
Indeed, soon after the birth of Einstein's theory of gravity, Cartan and Eisenhart revealed \cite{Cartan,Eisenhart1928} the possibility of two geometric approaches to nonrelativistic spacetimes and trajectories: (i) the ``intrinsic'' approach of Cartan and (ii) the ``ambient'' approach of Eisenhart.

On the one hand, Cartan advocated \cite{Cartan} that the notion of parallel transport is sufficient for a geometric reformulation of the equivalence principle, relativistic or not.\footnote{For instance, the trajectory of a freely falling observer in a gravity field is always described geometrically as an affine geodesic, in the sense of a curve in spacetime with tangent vector parallel transported along itself.} He thereby succeeded at geometrizing nonrelativistic spacetimes by defining them as manifolds endowed with absolute time and space (replacing the metric structure of relativistic spacetime), 
together with a compatible affine connection (so that parallelism tells matter how to move in spacetime) later dubbed ``Galilean'' connection.
On the other hand, Eisenhart proved \cite{Eisenhart1928} that the dynamical trajectories of nonrelativistic mechanics can always be lifted to geodesics of a specific relativistic spacetime of one dimension more possessing a null parallel vector field.
This class of spacetimes was discussed earlier by Brinkmann in a different context \cite{Brinkmann} and later received the interpretation of gravitational waves with parallel rays.
Conversely, nonrelativistic trajectories are obtained as the projection of geodesics of these relativistic spacetimes along these rays.
This correspondence between relativistic and nonrelativistic structures is also valid at the level of spacetime geometry so that the ambient approach turns out to be deeply related to the intrinsic approach of Cartan. More precisely, the quotient manifold of parallel rays of the relativistic spacetime is a submanifold of codimension-one that inherits a structure of nonrelativistic spacetime \cite{Duval:1984cj}.

The important lesson that one can draw from these seminal works is that, although nonrelativistic structures are usually not addressed in geometric terms and are often understood as mere limits of relativistic structures, on one side both structures can be defined geometrically and can live on their own, on the other side nonrelativistic structures can always be embedded inside relativistic structures thereby shedding new lights on the former.
The ambient approach has proved to offer a useful and fresh viewpoint on nonrelativistic physics.
Somewhat curious nonrelativistic features may acquire enlightening interpretations once they are translated into standard relativistic terms (for instance, nongeometric projective representations of the Galilei group arise from unitary representations of the Poincar\'e group)
and, vice versa, the null dimensional reduction often provides simple explanations regarding properties of some various gravitational waves (such as their geodesic completeness, their superposition principle, their field equations, \textit{etc}.).

The present paper is devoted to the geometric description of nonrelativistic particles (classical and first-quantized) in the ambient approach, for a class of relativistic spacetimes extending the one considered in \cite{Eisenhart1928,Duval:1984cj,Duval:1990hj}. 
More precisely, it is generalized to
ambient spacetimes admitting a hypersurface-orthogonal null Killing vector field \cite{Lichnerowicz,Julia:1994bs,Minguzzi:2006gq}. We present a new geometric definition of this extended class of spacetimes, more adapted to the description of particles. We emphasize the physical relevance of this class by developing in detail their interpretation as gravitational waves, by discussing some of their exceptional properties and by exhibiting interesting examples that appeared in the literature (such as Kaigorodov solutions, Schr\"odinger manifolds, \textit{etc}.) and that do not belong to the class initially considered  in \cite{Eisenhart1928,Duval:1984cj}.
In order to comment on this last point and to present further motivations, let us briefly sketch
the history of the ambient approach to nonrelativistic structures.

The work of Eisenhart \cite{Eisenhart1928} did not attract much attention from theoretical physicists for several decades, with the remarkable exception of Lichnerowicz \cite{Lichnerowicz}, who generalized the theorem of Eisenhart
to the above-mentioned class of spacetimes in the 1950s (but without providing any hint of their geometric or physical interpretation).
A reason might be that the surge of interest of relativists and field theorists for nonrelativistic mathematical structures only started in the late 1960s. In a sense, the field-theoretical analogue of the ambient approach is
Dirac's light-cone formalism \cite{Dirac:1949cp}. It was its development that indirectly led to the discovery of the group-theoretical avatars of the ambient approach, such as the embedding of nonrelativistic symmetry groups (\eg the Bargmann \cite{Bargmann:1954gh} and Schr\"odinger \cite{Niederer:1972zz} groups) inside their relativistic higher-dimensional counterparts (the Poincar\'e \cite{Gomis:1978fn} and conformal \cite{Sorba} groups, respectively) or the relation between the corresponding wave equations on flat spaces \cite{Gomis:1980jv}.
One may notice that it took almost seven decades before the respective approaches of Cartan and Eisenhart to nonrelativistic curved spacetimes and trajectories were unified in \cite{Duval:1984cj,Duval:1990hj}.
Actually, the authors of \cite{Duval:1984cj} independently rediscovered the results of Eisenhart \cite{Eisenhart1928} and generalized the ambient approach to gravity and to first-quantized particles.
Because of these historical detours, the embedding procedure is sometimes referred to as ``Bargmann'' framework \cite{Duval:1984cj} or as  ``Eisenhart'' lift \cite{Duval:1990hj}.
Since then, this formalism was successfully applied to a large variety of nonrelativistic problems, such as
Chern-Simons electrodynamics \cite{Duval:1994pw}, fluid dynamics \cite{Hassaine:1999hn}, Newton-Hooke cosmology \cite{Gibbons:2003rv}, Schr\"odinger symmetry \cite{Duval:2008jg,Duval:2009vt}, Kohn's theorem \cite{Gibbons:2010fb}, \textit{etc}. 
The ambient approach to gravity was extended in \cite{Julia:1994bs} precisely for the class of ambient spacetimes considered by Lichnerowicz \cite{Lichnerowicz} (this exact correspondence was observed in \cite{Minguzzi}).
This procedure of projecting along the null rays was called ``null'' \cite{Julia:1994bs} (or ``lightlike'' \cite{Minguzzi}) dimensional reduction since it was addressed as the counterpart of the ``spacelike'' dimensional reduction of Kaluza and Klein or the ``timelike'' dimensional reduction for stationary spacetimes.

More recently, the possibility of applying the techniques of holography to nonrelativistic systems \cite{Hartnoll:2008vx,Son:2008ye} again triggered a large wave of interest for the geometric approaches to nonrelativistic symmetries in the communities of relativists and field theorists.
A suggestive idea that quickly arose \cite{Goldberger:2008vg} was that the ambient approach might also apply to the holographic duality sketched in \cite{Son:2008ye}; in such case the correspondence would effectively reduce to a holographic duality where both sides (bulk and boundary) are nonrelativistic theories. Very recently, a higher-spin gravity dual to the unitary Fermi gas was proposed along similar lines \cite{Bekaert:2011cu}.
In these works, the background bulk geometry is asymptotically either an anti de Sitter or Schr\"odinger spacetime.
Such gravitational waves do not belong to the class of \cite{Eisenhart1928,Duval:1984cj,Duval:1990hj} but they do belong to
the one of \cite{Lichnerowicz,Julia:1994bs,Minguzzi:2006gq}; this further motivates the present study of this wider class.
Another motivation is that plane-fronted waves with parallel rays cannot be black (in the sense of possessing an event horizon)
while the extended class does contain black gravitational waves \cite{Hubeny:2003ug}.

The plan of the paper is as follows:
~\\In section \ref{Nonrelativistic}, after stating our notations and conventions (sec. \ref{Notations}), we review the results of \cite{Eisenhart1928,Lichnerowicz} first in the Lagrangian formalism (sec. \ref{Nonrelativistic Lagrangian} and \ref{Ambient Lagrangian}) and then from a Hamiltonian perspective (sec. \ref{Nonrelativistic Hamiltonian}) in order to motivate the class of spacetimes (called ``Platonic waves'') we will be interested in. Sections \ref{Heuristics} and \ref{Allegory} are dedicated to an illustration of the embedding of nonrelativistic physics inside relativistic spacetimes using the analogy proposed in \cite{Minguzzi} with Plato's allegory of the cave. The ambient approach is then applied to first-quantized particles in section \ref{Schrodinger} where Schr\"odinger equation is derived from Klein-Gordon equation. In section \ref{Brinkmann}, we focus on the whole class of gravitational waves (\ie  spacetimes admitting a null hypersurface-orthogonal vector field ) and show how they are endowed with a nonrelativistic absolute time (sec. \ref{GW structure}). We use this larger class to introduce a preferred set of coordinates (Brinkmann coordinates) in section \ref{Plato Brinkmann} as well as some terminology in sec. \ref{Platonic screens}. We next focus on Platonic waves by first considering a subclass (the one originally used in \cite{Eisenhart1928}), namely spacetimes admitting a null parallel vector field (dubbed ``Bargmann-Eisenhart waves'' in the following) in section \ref{Bargmann-Eisenhart spacetimes} and show how these waves are endowed with a full nonrelativistic structure \ie  an absolute time and an absolute space. Some properties and examples of Bargmann-Eisenhart waves are also discussed. We first provide our definition of Platonic waves (section \ref{Platonic}) as conformal Bargmann-Eisenhart waves with preserved null vector and discuss some of their properties. We then show the equivalence between Platonic waves and the class (studied in \cite{Julia:1994bs}) of gravitational waves whose hypersurface-orthogonal vector field is also Killing and make use of this definition to show that Platonic waves are the most general class of spacetimes inducing a nonrelativistic structure on its space of rays (section \ref{Julia-Nicolai}). Platonic waves are also shown to constitute a subset of (degenerate)-Kundt spacetimes in section \ref{Plato Kundt} and some physically relevant examples are discussed in sec. \ref{Miscellaneous}. We then make use of our definition of Platonic waves in order to show some results relative to their global and causal properties (section \ref{Global}) and to their curvature scalar invariants in section \ref{Further}.

\section{Nonrelativistic dynamical trajectories as geodesic motions}
\label{Nonrelativistic}
In this section, we start by introducing our notations and conventions. Then, we present the old results of  Eisenhart \cite{Eisenhart1928} and Lichnerowicz \cite{Lichnerowicz},
firstly, by reviewing the suggestive analogy proposed by Minguzzi between the null dimensional reduction and the allegory of the cave, secondly,
by motivating the form of the ambient metrics as an extension of some class of nonrelativistic Lagrangians 
and, thirdly, by checking explicitly that the null dimensional reduction of the geodesic equations for a 
specific class of spacetimes in $D$ dimensions boils down to the Euler-Lagrange equations of some 
holonomic dynamical systems of $d=D-2$ degrees of freedom.
However, this direct check in the Lagrangian framework (similar to the original proofs \cite{Eisenhart1928,Lichnerowicz}) is slightly cumbersome and partially obscures the simple mechanism behind the Eisenhart-Lichnerowicz theorem.
On the contrary, in the Hamiltonian formulation
this mechanism becomes 
more transparent. Since the Hamiltonian version seems not to have been discussed in detail yet in the literature, it is presented in the last subsection.

\subsection{Notations and conventions}
\label{Notations}
We will use the ``mostly plus'' convention for the signature of Lorentzian spacetimes. The nonrelativistic spacetime will be a manifold of dimension $n$ foliated by spatial hypersurfaces which are Riemannian manifolds of dimension $d=n-1$.
This manifold will be embedded inside an ambient relativistic spacetime of dimension $D=n+1$.
Minuscule greek indices $\mu,\nu,...$ will denote ``world'' (holonomic) ambient indices while
minuscule latin indices $a,b,...$ will denote ``tangent'' (anholonomic) ambient indices, both taking $D=n+1$ values ($0,1,2,\ldots,D-1$).
Minuscule latin indices as $i,j,...$ will denote (world or tangent) spatial indices taking $d=n-1$ values ($1,2,\ldots,d$).
When it will be pertinent, one introduces the Cartesian coordinates $\vec x=(z,\vec y)$ on Euclidean space ${\mathbb R}^d$.

\subsection{Basic heuristics of the ambient approach}
\label{Heuristics}
Before introducing the technical details of the null dimensional reduction, the key ideas will be presented pictorially by
pursuing the entertaining analogy proposed by Minguzzi between the ambient approach and the allegory of the cave \cite{Minguzzi}.

The allegory of the cave was presented by Plato in his celebrated work ``The Republic'' as an illustration of his theory of Forms \cite{Plato}.
Prisoners are chained in the middle of a cave. They face a blank wall; behind them is a fire.
They watch shadows projected on the wall in front of them by objects which move behind them and 
which they cannot see. In the allegory,
the two-dimensional shadows represent material phenomena that can be perceived while
the three-dimensional objects correspond to Plato's ideal Forms. According to Platonism, the ultimate reality is the world of Forms (3D objects), while Phenomena (2D images) are mere illusions because of the incomplete knowledge of mankind (prisoners).
Leaving aside these philosophical views and focusing on our topic, the allegory of the cave provides an ancient example of ``lightlike'' dimensional reduction where objects are projected on a codimension-one manifold along light rays.\footnote{In a sense, linear perspective in graphical arts
is an even simpler instance of ``lightlike'' dimensional reduction, where three-dimensional objects are represented on a two-dimensional surface via projection along visual light rays. However, this example is not as useful for illustrating our purpose because linear perspective is static
while time plays a crucial role in the ambient construction.}
The analogy between the allegory of the cave and the ambient approach is even closer (Table \ref{table1}):
consider an ambient spacetime (playing the role of the cave in the allegory) on which a gravitational wave propagates and to which corresponds a congruence of graviton worldlines (replacing the light rays emitted by the fire). Physicists detect the corresponding gravitons on a screen (the wall where photons are projected in the allegory).\footnote{The switch from the gravitational wave to the graviton description is simply understood by applying the standard rules of translation (between wave and particle language) from geometric optics 
where the propagation of wavefronts is equivalently described by its orthogonal rays, which can be interpreted as worldlines.}
This projection of ambient events on the screen along gravitational rays is the most concrete way of formulating the null dimensional reduction considered in this paper.
The main lesson from the ambient approach is that the relativistic spacetime and the particle trajectories appear nonrelativistic when read on the screen. In this sense, nonrelativistic structures are mere shadows of relativistic ones.

In order to present the heuristics behind this mathematical fact,
notice that the screen registers the following events: absorption or emission of a graviton by the screen.
These events are encoded via the position on the screen and the instant of the intersection.
The description of the screen worldvolume (\ie the time evolution of the screen) via these coordinates already suggests that the former might be endowed with a natural structure of (codimension-one) spacetime. What is more remarkable is that this structure is nonrelativistic and that the shadows on the screen from ambient geodesics have a natural interpretation as dynamical trajectories of nonrelativistic particles.

\begin{table}
\begin{center}
\begin{tabular}{
|c|c|}
 \hline
Allegory of the cave & Ambient approach\\
\hline\hline
Cave & Ambient spacetime\\\hline
Wall & Screen \\\hline
Light rays & Graviton worldlines\\\hline
Shadows & Nonrelativistic physics\\\hline
\end{tabular}
\end{center}
\caption{Analogy: 
Allegory of the cave / Ambient approach\label{table1}}
\end{table}

\subsection{Nonrelativistic Lagrangian}\label{Nonrelativistic Lagrangian}

Consider a smooth manifold with coordinates $(t,x^i)$ and the most general Lagrangian that is a polynomial of degree two in the velocities $\dot{x}^i=dx^i/dt$:
\bea
L(t,x,\dot{x})&=&\frac12\,\bg_{ij}\pl t,x\pr \dot{x}^i \dot{x}^j
+\bA_i\pl t,x\pr \dot{x}^i-\bar V\pl t,x\pr\,  \label{lagrang}
\eea
where $\bg_{ij}$ is sometimes called the mass matrix. In order to avoid ghosts and constraints, we require the kinetic term $\frac12\,\bg_{ij}\pl t,x^i\pr \dot{x}^i \dot{x}^j$ to be a positive-definite quadratic form in the velocities.
A dynamical system described by \eqref{lagrang}
can always be interpreted as describing the motion of a charged particle minimally coupled to an electromagnetic field through the vector potential $\bA_i$ and the scalar potential $\bar V$, called ``effective'' potential in the following, and moving on a Riemannian manifold with metric $\bg_{ij}$.

Leaving aside this interpretation, this class of Lagrangians corresponds to the most general holonomic dynamical system obeying d'Alembert's principle with external forces $F_i=\bar F_{ij}\dx^j+\bar F_i$ at most linear in the velocity satisfying the two further requirements: the linear part $\bar F_{ij}\dx^j$ in the velocity of the external force does not develop any power ($\bar F_{ij}\dx^i\dx^j=0\Leftrightarrow \bar F_{(ij)}=0$ \footnote{Curved (respectively, square) brackets over a set of indices denote complete (anti)symmetrization over all these indices, with weight one, \textit{i.e.} $S^{(\mu_1\ldots \mu_r)}=S^{\mu_1\ldots \mu_r}$ and
$A^{[\mu_1\ldots \mu_r]}=A^{\mu_1\ldots \mu_r}$ respectively for $S$ and $A$ totally (anti)symmetric tensors.}) and derives from a vector potential ($\bar F_{ij}=2\partial_{[i}\bar A_{j]}$) while the part independent of the velocity derives from a scalar potential ($\bar F_i=-\p_t\bar A_i-\partial_i \bar V$). The vector and effective potentials may depend on time. The Lorentz force is indeed the perfect example of such an external force. 
For later purpose, let us emphasize that the holonomic coordinates $x^i$ of a given holonomic system are only defined up to a reparametrization
\bea
x^i&\to& x^{\prime i}=x^{\prime i}(t,x) \label{repar}\\
t&\to& t'=t\nn
\eea
which preserves the general form of \eqref{lagrang}, but redefines the various coefficients $\bg_{ij}$, $\bA_i$ and $\bar V$.

Let us emphasize that the gift of the Lagrangian \eqref{lagrang} defines a nonrelativistic spatial metric on the manifold labeled by the coordinates $(t,x^i)$. In other words, the mass matrix $\g_{ij}$, being positive definite, provides a collection of rulers at any event. As the notion of a nonrelativistic spacetime necessitates absolute rulers \textit{and} clocks, this motivates the introduction of a collection of clocks, equivalent to the gift of a function $\Omega(t,x)>0$ specifying the unit of time at each point of spacetime. The lapse $d\tau'=m\,d\tau$ of local time $\tau'$ measured by the local clock (along a trajectory) corresponding to the lapse $dt$ of absolute time $t$ is: 
\bea
d\tau'=\Omega(t,x)\,dt=m\,d\tau\label{dtau'}, 
\eea
where the constant $m$ is introduced by analogy with affine parameters (which are also defined up to a multiplicative constant $\tau'=m\,\tau$) and will acquire soon the interpretation of a nonrelativistic mass.

Since our goal is to relate the Lagrangian \eqref{lagrang} to the geodesic equation for some spacetime, let us stress the similarities and differences of such an action principle with the quadratic action principle for a geodesic. Suggestively, one can rewrite the action
\bea
S[\,x^i\,]=m\int L(t,x,\dot{x})\,dt\,\label{action}
\eea
corresponding to the nonrelativistic Lagrangian \eqref{lagrang} in terms of the local time along the trajectory as
\bea\label{paraction}
S[\,x^i\,]=\int{ \Omega\left(\half\, \bg_{ij} \frac{dx^i}{d\tau}\frac{dx^j}{d\tau}
+\bA_i\frac{dx^i}{d\tau}\frac{dt}{d\tau}-\bar V \frac{dt}{d\tau}\frac{dt}{d\tau}\right)\,d\tau\,}.
\eea
where eq.\eqref{dtau'} has been used. With the classical action \eqref{action} being defined up to a multiplicative constant, the factor $m$ has been introduced for later purposes. 
Notice that the case $m=0$ is special and corresponds to nondynamical trajectories in the sense that eq.\eqref{dtau'} implies $dt=0$ and so the curve $\big ( t,x^i\pl \tau\pr\big )$ is at fixed $t$. Moreover, the action \eqref{paraction} becomes $S[\,x^i\,]=\half\int{ \Omega\, \bg_{ij} \frac{dx^i}{d\tau}\frac{dx^j}{d\tau}d\tau}$ which has the form of a quadratic geodesic action for the metric $g_{ij}=\Om\, \g_{ij}$.  
 
The action \eqref{paraction} looks like the quadratic action for a geodesic in the spacetime described by the line element:
\bea
ds_{(n)}^2&=&\Omega\left(\bg_{ij} dx^i dx^j
+2\,\bA_i\,dx^idt-2\,\bar V dt^2\right)\nonumber\\
&=&g_{ij} dx^i dx^j
+2\,A_i\,dx^idt-2\,V dt^2\,.\label{rhsmet}
\eea
However, an important discrepancy between \eqref{paraction} and the action principle for a geodesic corresponding to the line element \eqref{rhsmet} is that the parameter $\tau$ is not an affine parameter since its normalization is not defined in terms of the metric defined by \eqref{rhsmet} but simply
as
\bea
\Omega\frac{dt}{d\tau}=m\,.\label{normali}
\eea
Although the right-hand side \eqref{rhsmet} can na\"ively be interpreted as a line element on the nonrelativistic $n$-dimensional spacetime, this metric has actually no definite signature since there is no \textit{a priori} sign constraint on the potential $V$ (which might even be vanishing).
Nevertheless, the gift of a Lagrangian of degree two in the velocities and of a time unit is equivalent to the gift of an indefinite line element of spacetime. However, a nonrelativistic spacetime has a somewhat weaker structure: it is rather defined only by the clocks $\Omega(t,x)\,dt$ and by the rulers encoded in the spatial metric $d\ell^2=g_{ij}(t,x) dx^i dx^j$ on the spatial leaves $t$=const. 

In order to lift the dynamical trajectories ($m\neq 0$) to geodesics of an ambient spacetime, the crucial ingredient is to add the value of the action as an extra coordinate. More precisely, we introduce a coordinate $u$ proportional to the action and to the local time $\tau$ such that
the infinitesimal variation of the action \eqref{action} along a trajectory is equal to
\bea du=-L\,dt-\frac{M^2}{m}\, d\tau\label{du}. \eea 
The minus sign and normalization have been chosen for later convenience.
By making use of the relations \eqref{lagrang} and \eqref{du}, the line element \eqref{rhsmet} is equal to:  
\bea
\Omega\left(\bg_{ij} dx^i dx^j
+2\,\bA_i\,dx^idt-2\,\bar V dt^2\right)=-2\Om\, dt du-2M^2d\tau^2\label{dtdudtau}. 
\eea
The main idea behind the Eisenhart lift (in Lagrangian terms) is to make use of \eqref{normali} in order to reinterpret this relation as expressing the fact that $\tau$ is an affine parameter along a geodesic in an ambient spacetime
of coordinates $x^\mu\equiv(u,t,x^i)$ and suitable metric $g_{\mu \nu}$. More precisely, we want to rewrite \eqref{dtdudtau} as the relation $g_{\mu \nu}dx^\mu dx^\nu=-M^2d\tau^2$ where the constant $\left\vert M^2 \right\vert$ stands for the ambient velocity norm squared. We will check that eq. \eqref{normali} simply arises as an equation of motion.
We should stress that there is a large ambiguity in reading off the ambient metric from \eqref{dtdudtau} when the geodesics are not lightlike $\pl M^2\neq0\pr$. 
More precisely, the relation
\eqref{dtdudtau} can be rewritten as a normalization condition for the affine parameter $\tau$:
\bea
 \Omega\pl t,x\pr \big[2\,dt\left(du +\bA_i\pl t,x\pr  dx^i-\bU\pl t,x\pr  dt\right)+\bg_{ij}\pl t,x\pr  dx^i dx^j
\big]= -M^2\, d\tau^2\,\label{dtduetc}
\eea
if we define
\bea
\bar U=\bar V -\half\,\frac{M^2}{m^2}\,\Omega\,.\label{UbarVbar}
\eea
In order to distinguish them, the potential $\bar V$ will be referred to as \textit{effective potential} while the term \textit{scalar potential} will be reserved to designate $\bU$. 
If the geodesic is lightlike, then $M^2=0$ and thus $\bar U=\bar V$. 
The left-hand side of \eqref{dtduetc} can be interpreted as the ambient line element 
\bea
ds^2&=&\Om\pl t,x\pr  \crl2\,dt\pl  du+\bA_i\pl t,x\pr 
 dx^i-\bU\pl t,x\pr  dt\pr \,+\,\g_{ij}\pl t,x\pr  dx^i dx^j\crr\label{Llinel}\\
&=& 2\,\Om\pl t,x\pr dt du+2\,A_i\pl t,x\pr 
 dt dx^i-2\,U\pl t,x\pr  dt^2\,+\,g_{ij}\pl t,x\pr dx^i dx^j\nonumber
\eea
The ambient metric $g$ is conformally equivalent to the metric $\bg$ with line element
\bea
d\bar s^2=2\,dt\pl  du+\bA_i\pl t,x\pr 
 dx^i-\bU\pl t,x\pr  dt\pr \,+\,\g_{ij}\pl t,x\pr  dx^i dx^j\,\label{Elinel}
\eea
in the sense that  \bea
g_{\alpha  \beta}=\Omega\dep\bg_{\alpha\beta}\,. \label{conformalBargmann}
\eea
Line elements of the form \eqref{Elinel} were considered by Eisenhart in \cite{Eisenhart1928}, while Lichnerowicz \cite{Lichnerowicz}
introduced the line element \eqref{Llinel}, but none of them provided an explanation for their choice of metrics or a reason why the null dimensional reduction precisely works for this large class of metrics.
The chain of arguments presented in this subsection is intended as a plausible line of reasoning leading to this choice.

\noindent\textbf{Remark 1:} Given an effective potential $\bar V$, eq.\eqref{UbarVbar} shows that to any choice of time unit $\Om$ corresponds distinct ambient metrics \eqref{Llinel}. Therefore, to a given Lagrangian system corresponds an infinite class of relativistic spacetimes not considered in \cite{Eisenhart1928}.  

\noindent\textbf{Remark 2:} Let us remind the reader that two Lagrangians $L$ and $L'$ are said to be equivalent if the actions differ by a total derivative, 
$L'=L+\frac{df}{dt}$, since their Euler-Lagrange equations are identical. In terms of the potentials, this is equivalent to a gauge transformation $\bar A'_i=\bar A_i+\partial_i f$ and $\bar V'=\bar V-\partial_t f$. From the point of view of the action, this means they differ by a boundary term, essentially equal to the variation of the function $f$. The interpretation of the variation of $u$ as linear in the variation of the action along the trajectory suggests that the previous equivalence corresponds to the reparametrizations $u'= u+f(t,x)$.
One can indeed check that the form \eqref{Llinel} of the line element is preserved by this coordinate transformation, up to a gauge transformation of the potentials. 

\subsection{Ambient Lagrangian}
\label{Ambient Lagrangian}

Consider now the action principle $S\crl x^\mu\crr=\int{ \Lag \,d\tau}$ for the geodesics parametrized by the affine parameter $\tau$, on the ambient spacetime with line element \eqref{Llinel}, where the quadratic Lagrangian reads
\bea\label{quadrL}
\Lag\crl x^\mu,\frac{d x^\nu}{d\tau}\crr=\half \,g_{\alpha\beta}\pl t,x\pr\dxata\dxbta.
\eea
The affine parameter $\tau$ is defined by the affine parametrization constraint $\Lag=-\frac{M^2}{2}$, which is nothing but \eqref{dtduetc}.
The equations of motion read
\bea
\mbox{for } u:&&\der{}\pl\Om\der{t}\pr=0\label{eomcbu1}\\
\mbox{for }t:&&\der{}\crl\Omega\pl\der{u}-2\bU\der{t}+\bA_i\der{x^i}\pr\crr=-\frac{M^2}{2\Om}\p_t\Om\nn\\&&+\Om\pl-\p_t \bU\pl\der{t}\pr^2+\p_t\bA_i\der{t}\der{x^i}+\half
\p_t \g_{ij}\der{x^i}\der{x^j}\pr\label{eomcbt1}\\
\mbox{for }x^i:&&\der{}\crl\Om\pl \g_{ij}\der{x^j}+ \bA_i\der{t}\pr\crr=-\frac{M^2}{2\Om}\p_i\Om\nn\\&&+\Om\pl-\p_i \bU\pl\der{t}\pr^2+\p_i\bA_j\der{t}\der{x^j}+\half\p_i\g_{kl}\der{x^k}
\der{x^l}\pr\label{eomcbx1}
\eea
where the affine parametrization constraint $\Lag=-\frac{M^2}{2}$ has been used to simplify \eqref{eomcbt1}-\eqref{eomcbx1}.
We can solve eq.(\ref{eomcbu1}) in the form of \eqref{normali} where $m$ is now interpreted as a constant of motion, $\der{m}=0$. This conservation law comes from the fact that the Lagrangian \eqref{quadrL} does not depend on $u$.
Thus the condition \eqref{normali} is obtained as an equation of motion. Two cases must be distinguished: $m=0$ and $m\neq0$.
The particular case $m=0$ corresponds to the geodesics that entirely belong to a given hypersurface $t$=const since $dt/d\tau=0$. Contrarily to the generic case $m\neq 0$, these curves have no interpretation as dynamical trajectories: they are the rays of the congruence. 
\vspace{1mm}

\noindent$\bullet$ \underline{$m=0$, $M^2=0$ (null rays):} 
If the geodesic is lightlike then the affine parametrization constraint \eqref{dtduetc} with $dt=0$
implies that $dx^i/d\tau=0$. The latter equation together with $dt/d\tau=0$ inserted into the equation of motion \eqref{eomcbt1}
imply that $du/d\tau$=const, since $\Omega(t,x)$=const. In conclusion, the lightlike geodesics belonging to a hypersurface of constant $t$ are curves with $x^i$ constant and with $u$ as an affine parameter. These are the graviton worldlines defining the gravitational wave. As one can see, they generate the 
hypersurfaces $t$=const which are called ``wavefront worldvolumes''. A locus $u=f\pl t,x\pr$ defines a screen of detection/emission. 

\vspace{1mm}

\noindent$\bullet$ \underline{$m=0$, $M^2< 0$ (spatial trajectories):} One can check that the spacelike geodesics are at the same time geodesics $x^\mu\pl\tau\pr$ of the $D$-dimensional ambient spacetime
 and project onto spatial geodesics $x^i\pl\tau\pr$ of the metric $g_{ij}=\Omega\,\bar g_{ij}$. This can be seen by checking that eq.\eqref{eomcbx1}
with $dt/d\tau=0$ is equivalent to the geodesic equation for the metric $g_{ij}$ and the affine parametrization constraint reads ${\cal L}=\half \,g_{ij}\dxita\dxjta=-\frac{M^2}{2}$.
In this sense, the wavefront worldvolumes $t=$const are totally geodesic submanifolds of the ambient spacetime.

\vspace{1mm}

\noindent$\bullet$ \underline{$m\neq 0$ (dynamical trajectories):} In the generic case $m\neq 0$, one can reexpress eqs(\ref{eomcbt1})-(\ref{eomcbx1}) as:
\bea
&&\ddu-\p_t\bU-2\p_i\bU\dx^i+\p_i\bA_j\dx^i\dx^j+\bA_i\ddx^i-\half\p_t\g_{ij}\dx^i\dx^j+\frac{M^2}{2m^2}\p_t\Om=0\label{eomt31}\\
&&\ddx^m+\bar \Gamma^m_{lj}\dx^l\dx^j+\g^{km}\crl\dx^i\pl\p_t\g_{ki}+\p_i\bA_k-\p_k\bA_i\pr+\p_k\bU+\p_t\bA_k\crr\nn\\&&+\frac{M^2}{2m^2}\p_k\Om \g^{km}=0\label{eomx31}
\eea
We can put eq.(\ref{eomx31}) in the form 
of the Euler-Lagrange equation for the original Lagrangian \eqref{lagrang}
\bea
\ddx^m+\bar \Gamma^m_{lj}\dx^l\dx^j+\g^{km}\crl\pl\p_t\g_{ki}+\bar F_{ik}\pr\dx^i-\bar E_k\crr=0
\eea
where we introduced the spatial Levi-Civita connection $\bar \Gamma^m_{lj}$, the magnetic field strength $\bar F_{ik}=\p_i\bA_k-\p_k\bA_i$ and the electric field
$\bar E_k=-\p_k\bar V-\p_t\bA_k$
together with the definition
\eqref{UbarVbar}. Moreover, it can be checked that eq.\eqref{eomt31} is compatible with the expression for $\dot u$ coming from the affine parametrization constraint \eqref{dtduetc}.

This completes the explicit check that the geodesics with $m\neq 0$ for the ambient spacetime \eqref{Llinel} correspond to dynamical trajectories for the 
Lagrangian \eqref{lagrang} in terms of the coordinates $x^i$ and $t$ so that the Eisenhart-Lichnerowicz theorem can now be formulated as:
\bthm{Eisenhart-Lichnerowicz \cite{Eisenhart1928,Lichnerowicz}}{The null dimensional reduction along the direction $u$ of the affine geodesic equation for a curve $x^\mu\pl\tau\pr=\pl u\pl\tau\pr,t\pl \tau\pr,x^i\pl\tau\pr\pr$ parameterised by the affine parameter $\tau$, satisfying $\frac{dt}{d\tau}\neq0$ and $g_{\mu\nu}\frac{dx^\mu\pl\tau\pr}{d\tau}\frac{dx^\nu\pl\tau\pr}{d\tau}=-M^2$ on a manifold endowed with the metric 
\bea
ds^2&=&\Om\pl t,x\pr  \crl2\,dt\pl  du+\bA_i\pl t,x\pr 
 dx^i-\bU\pl t,x\pr  dt\pr \,+\,\g_{ij}\pl t,x\pr  dx^i dx^j\crr\nonumber
\eea
reduces to the Euler-Lagrange equations of the holonomic dynamical system characterised by the Lagrangian
\bea
L(t,x,\dot{x})&=&\frac12\,\bg_{ij}\pl t,x\pr \dot{x}^i \dot{x}^j
+\bA_i\pl t,x\pr \dot{x}^i-\bar V\pl t,x\pr\,  \nn
\eea
where the effective potential $\bar V$ reads $\bar V=\bar U +\half\,\frac{M^2}{m^2}\,\Omega$, with $m=\Om\frac{dt}{d\tau}$. 
 }
We remind the reader that the extra coordinate $u$ can be interpreted as
the value of the action evaluated along the trajectory.

\subsection{Hamiltonian perspective}\label{Nonrelativistic Hamiltonian}

The momenta corresponding to the Lagrangian \eqref{lagrang} are given by $p_i=\bg_{ij}\pl t,x\pr \dot{x}^j
+\bA_i\pl t,x\pr $. Thus the Hamiltonian reads
\bea
H(t,x^i,p_j)&=&\frac12\,\bg^{ij}\pl t,x\pr \Big( p_i-\bA_i\pl t,x\pr\Big)\Big( p_j-\bA_j\pl t,x\pr\Big) +\bar V\pl t,x\pr\,  \label{hamilton}
\eea
where $\bg^{ij}$ denotes the inverse of the metric $\bg_{ij}$. Obviously, this Hamiltonian function is the most general polynomial of degree two in the momenta with a positive-definite quadratic form as leading term.

The connection between the Hamiltonian action principles for the dynamical trajectories and for the ambient geodesics
will be manifest in the ``parametrized'' Hamiltonian formulation obtained from the Lagrangian formulation where $t(\tau)$ is taken as a dynamical degree of freedom.
The detailed Hamiltonian analysis\footnote{See \eg \cite{Henneaux} for more details on parametrized systems and their Hamiltonian constraints. Let us stress that, in the parametrized Hamiltonian formulation, the canonical Hamiltonian vanishes because of the time reparametrization invariance.} of such a system leads to the following action principle:
\bea\label{par1}
S[t,x^i,p_t,p_j,\lambda]=\int{\crl p_i\frac{dx^i}{d\tau}+p_t\,\frac{dt}{d\tau}-\lambda\Big( p_t+H\pl t,x^i,p_j\pr\Big)\crr}d\tau
\eea
where $p_t$ is the conjugate of the (now dynamical) variable $t$ while $\lambda$ is the Lagrange multiplier for the first-class\footnote{A single constraint is automatically first class.} constraint
$p_t+H=0$ corresponding to the reparametrization invariance of the parameter $\tau$. 
Solving the constraint as $p_t=-H$ inside the action gives the equivalent action principle
\bea
S[t,x^i,p_j]=\int{\crl\, p_i\frac{dx^i}{d\tau}\,-\,H\pl t,x^i,p_j\pr\frac{dt}{d\tau}\,\crr}d\tau
\eea
where the reparametrization invariance $\tau\to\tau'=\tau'(\tau)$ can be used to impose the gauge fixation $dt/d\tau=1$ in order to get the usual action principle $S[x^i,p_j]=\int{\crl p_i\dx^i-H\pl t,x^i,p_j\pr\crr}dt.$ 

Now let us consider the parametrized Hamiltonian formulation of a free relativistic particle of mass $M$ propagating on the ambient spacetime
with line element \eqref{Llinel} that arises from the Lagrangian $\mathcal L'=-M\sqrt{\left \vert g_{\alpha \beta}\frac{dx^\alpha}{d\tau}\frac{dx^\beta}{d\tau}\right \vert}$:
\bea
\mathcal{S}[x^\mu,p_\nu,\lambda]=\int{\crl p_\mu\frac{dx^\mu}{d\tau}-\frac{\lambda}{2}\,\Omega\pl p^2+M^2\pr\crr}d\tau\,,\label{action6}
\eea
with $p_u=\Omega\frac{dt}{d\tau}$ and $\lambda$ a Lagrange multiplier for the mass-shell constraint $p^2+M^2=0$ and where
\bea
p^2&=&g^{\mu\nu}p_\mu p_\nu\label{psqu}\\
&=&\Omega^{-1}\pl t,x\pr\Big[2\,p_tp_u+\,\bg^{ij}\pl t,x\pr \Big( p_i-\bA_i\pl t,x\pr p_u\Big)\Big( p_j-\bA_j\pl t,x\pr p_u\Big)\nonumber\\
&&\qquad \qquad \qquad \qquad\qquad \qquad\qquad \qquad\qquad \qquad\qquad \qquad
 +2\,\bar U\pl t,x\pr p_u^2\Big]\,.\nonumber
\eea
As one can see, the form of the inverse metric $g^{\mu\nu}$ can be characterized as the most general ambient inverse metric that is independent of $u$ and such that $g^{t\mu}\propto \delta^{u\mu}$. These two properties turn out to be the only two crucial ingredients in the null dimensional reduction of the Hamiltonian. This again provides a justification for the line element \eqref{Llinel}.

When $p_u\neq 0$, it turns out to be convenient to define 
\bea
\bar U=\bar V -\half\,\frac{M^2}{p_u^2}\,\Omega\,,\label{UbarVbar'}
\eea
because inserting \eqref{psqu}-\eqref{UbarVbar'} inside \eqref{action6} leads to a form of the action which is suggestively close to \eqref{par1}:
\bea
\mathcal{S}[x^\mu,p_\nu,\lambda]=\int d\tau\crl p_i\frac{dx^i}{d\tau}+p_t\frac{dt}{d\tau}+p_u\frac{du}{d\tau}-\lambda  \pl p_tp_u+\,{\cal H}(t,x^i,p_j,p_u)\pr\crr\label{actambham}\,,
\eea
with
\bea\label{actambham2}
{\cal H}(t,x^i,p_j,p_u)&=&\frac12\,\bg^{ij}\pl t,x\pr \Big( p_i-\bA_i\pl t,x\pr p_u\Big)\Big( p_j-\bA_j\pl t,x\pr p_u\Big)\nonumber\\ &&\qquad +\bar V\pl t,x\pr p_u^2 \,.
\eea
The form of this Hamiltonian is the most general function of $x^\mu$ and $p_\mu$ that is a homogeneous polynomial of degree two in the momenta \textit{and} independent of $u$ and $p_t$. It can be seen as the homogenization of the original Hamiltonian \eqref{hamilton}.

The main difference between the ambient action principle \eqref{actambham}-\eqref{actambham2} and the reduced action principle \eqref{hamilton}-\eqref{par1}
is the dependence on the conjugate pair of variable $u$ and $p_u$. The decisive observation is that, since there is no explicit dependence on the variable $u$ in the Hamiltonian \eqref{actambham2}, the conjugate momentum $p_u=\Omega\frac{dt}{d\tau}=m$ is a constant of motion. Therefore, it will not play any role in the Hamilton equations for the remaining variables which will thus be essentially the same as the original system. This proves the Eisenhart-Lichnerowicz theorem without the need for performing any tedious computation.
In Hamiltonian language, this theorem may be phrased simply as follows: the original system \eqref{hamilton}-\eqref{par1} can be seen as the symplectic reduction of the system \eqref{actambham}-\eqref{actambham2} through the addition of the extra constraint $p_u-m=0$, which is first-class since ${\cal H}$ is independent of $u$. In other words, the action principle \eqref{par1} is equivalent to the action principle
\bea
S[x^\mu,p_\nu,\lambda,\mu]&=&\int d\tau\Big[ p_i\frac{dx^i}{d\tau}+p_t\frac{dt}{d\tau}+p_u\frac{du}{d\tau}\nonumber\\
&&\qquad-\lambda  \pl p_tp_u+\,{\cal H}(t,x^i,p_j,p_u)\pr-\mu \big(p_u-m\big)\Big]\,,
\eea
where $\mu$ is a new Lagrange multiplier enforcing the constraint $p_u=m$.

Retrospectively, from the parametrized Hamiltonian perspective the main trick behind the ambient approach to dynamical trajectories is the homogenization of the constraint $p_t+H(t,x^i,p_j)=0$ to get a constraint $p_tp_u+{\cal H}(t,x^i,p_j,p_u)=0$ that is quadratic in the momenta, via the introduction of an auxiliary momentum coordinate. The resulting constraint is a nondegenerate quadratic polynomial in the momentum with Lorentzian signature and can therefore be interpreted as the mass-shell constraint $p^2+M^2=0$ of a free relativistic particle. There is an arbitrariness in such an identification which is reflected in the relation \eqref{UbarVbar'}.

As a side remark, one may notice that by dividing \eqref{actambham2} by $p_u$, one may see that the auxiliary momentum $p_u$ actually plays the role of a nonrelativistic mass (\eg the kinetic term of the ``light-cone Hamiltonian'' ${\cal H}/p_u$ is of the form $\vec p^2/2m$). 
This remark provides a nice interpretation of the action obtained from \eqref{actambham}
after solving the mass-shell constraint as $p_t=-{\cal H}/p_u$
and fixing the reparametrization invariance by $\tau=t$:
\bea S[x^i,u,p_j,p_u]=\int{\crl p_i\dx^i+p_u\du-\frac{{\cal H}\pl t,x^i,p_j,p_u\pr}{p_u}\crr}dt.\eea 
This interpretation of the auxiliary momentum $p_u$ as a nonrelativistic mass is standard when the ambient spacetime is Minkowski (or AdS) spacetime. In such cases, the ambient approach essentially coincides with the light-cone formalism, but a remarkable fact is that this setting actually generalizes smoothly to the much wider class of curved spacetimes with line element \eqref{Llinel} that will be motivated and described more geometrically in the following. 
\subsection{Gravitational waves and Plato's allegory}
\label{Allegory}
In order to understand better the heuristics behind the ambient approach,
let us describe the former spacetimes in more geometric terms, starting to sketch some technical details and motivating our future choices of terminology.

\begin{figure}[ht]
\centering
   \includegraphics[width=0.5\textwidth]{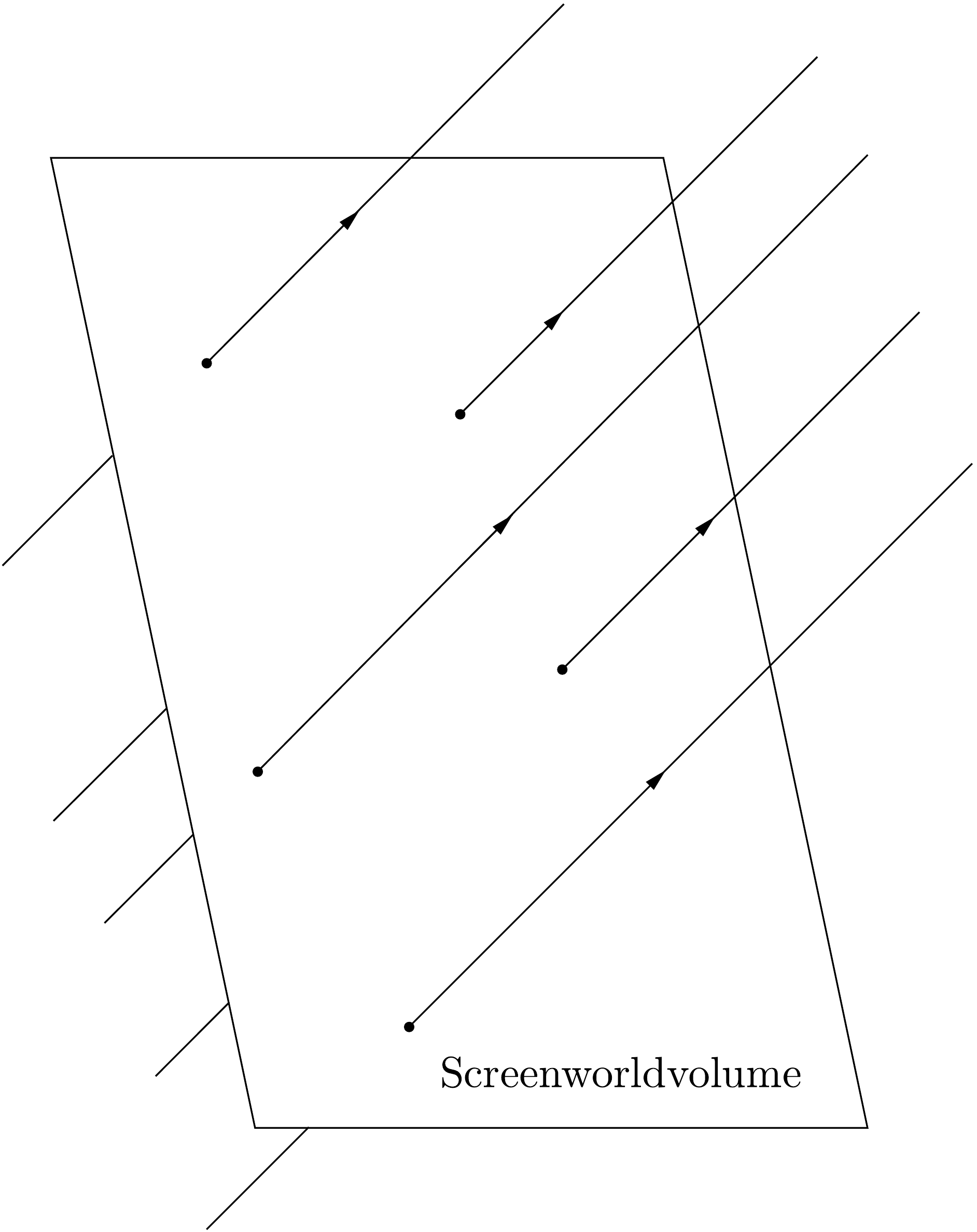}
  \caption{The screen worldvolume is transverse to the congruence of null rays. (In all figures, we will follow the standard spacetime diagram convention, \ie time flows from bottom to top and null directions are at 45$^\circ$.)}\label{fig1}
\end{figure}

\begin{figure}[ht]
\centering
   \includegraphics[width=0.7\textwidth]{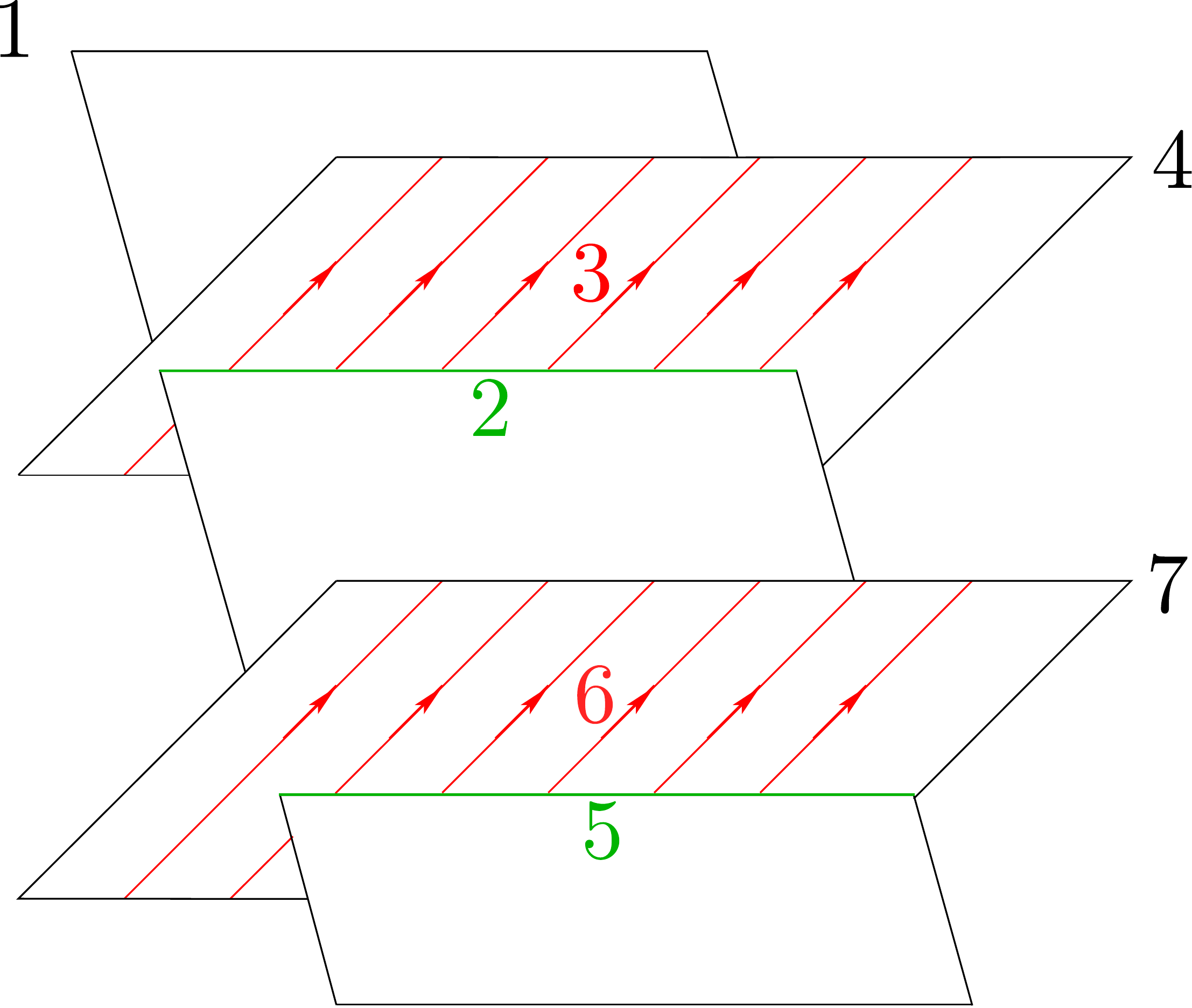}
  \caption{1. Screen worldvolume;
2. Screen at $t=t_1$;
3. Congruence of null geodesics generating the wavefront worldvolume $t=t_1$;
4. Wavefont worldvolume $t=t_1$;
5. Screen at $t=t_0$;
6. Congruence of null geodesics generating the wavefront worldvolume $t=t_0$;
7. Wavefont worldvolume $t=t_0$}\label{fig2}
\end{figure}

Consider the propagation of a gravitational wave in the ambient spacetime and a screen detecting the gravitons passing by.
In a spacetime diagram, the worldlines of gravitons are null rays, \textit{i.e.} they define a null geodesic congruence, and the registered events on the screen are
simply intersections between the screen worldvolume and the null rays.
So, technically, the screen worldvolume is a codimension-one hypersurface which is transverse to the congruence of null rays, in the sense that each ray intersects it only once (Fig. \ref{fig1}).
The events are encoded via the position on the screen and the instant of the intersection.
Heuristically, these coordinates on  the screen worldvolume already suggest that the former might be endowed with a natural structure of (codimension-one) spacetime.
In order to push the spacetime picture further, consider the screen at any given instant as a wavefront. From a spacetime point of view, the propagation of this wavefront
translates into the fact that null rays generate the corresponding wavefront worldvolume, each such hypersurface is labeled by the time of emission, the ``retarded'' time (Fig. \ref{fig2}).
The family of these wavefront worldvolumes provides a foliation of the ambient spacetime the leaves of which are orthogonal to the null rays.
Retrospectively, this provides a geometric definition for a gravitational wave as a foliated spacetime. The screen worldvolume can then be thought as a codimension-one hypersurface transverse to this foliation, such that the intersection between a leave and the screen worldvolume is precisely the instantaneous screen we started with.

The projection on the screen along rays maps the ambient spacetime on a codimension-one manifold endowed with a notion of time induced
from the foliation of the ambient spacetime: the retarded time.
If the relativistic structure (\textit{i.e.} the metric) of the ambient spacetime is preserved along the rays (\textit{i.e.} they are Killing orbits),
then it can induce a well-defined structure on the quotient space which can be represented as a screen worldvolume.
The remarkable fact is that this projection defines a nonrelativistic spacetime structure (\textit{i.e.} absolute rulers and clocks) on the screen worldvolume.\footnote{By construction, this structure does not depend on the specific choice of screen worldvolume, for instance two screens in relative motions would encode the same geometric data with respect to their rulers and clocks.}

\begin{figure}[ht]
\centering
   \includegraphics[width=0.7\textwidth]{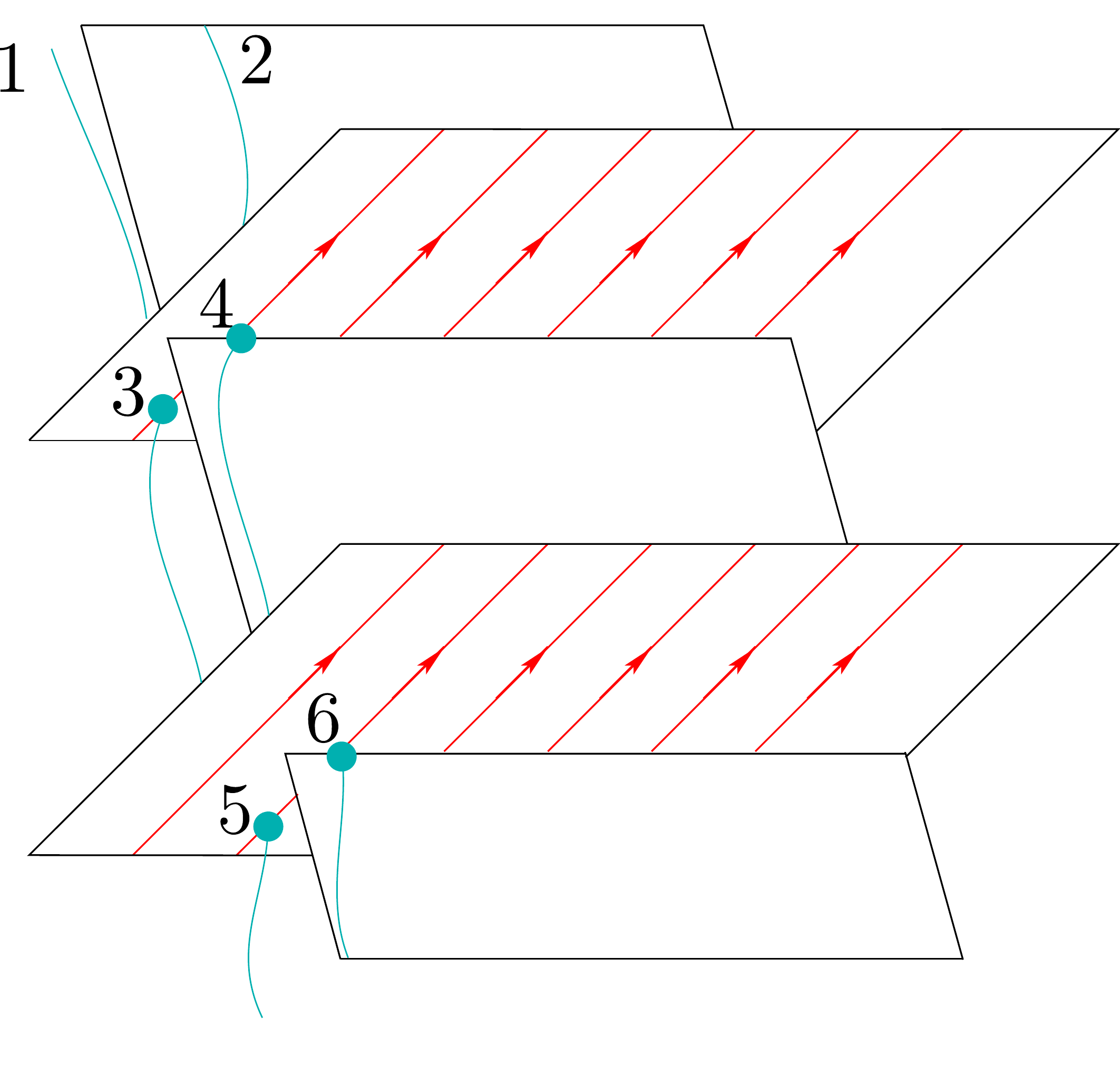}
  \caption{The Eisenhart lift,  1. Geodesic of ambient spacetime;
2. Shadow of the ambient geodesic on the screen worldvolume;
3. Emission of a graviton by the geodesic;
4. Detection on the screen at $t=t_1$; 
5. Emission of a graviton by the geodesic;
6. Detection on the screen at $t=t_0$;}\label{fig3}
\end{figure}

Actually, the induced line element on the screen worldvolume encodes more information than absolute clocks and rulers but is equivalent to the specification of a Lagrangian for a holonomic dynamical system.
Perhaps even more remarkable is that the projections of ambient geodesics on the screen have a natural interpretation as dynamical trajectories of nonrelativistic particles (Fig. \ref{fig3}). Furthermore, between the emission of a graviton by the geodesic and its detection on the screen,
the affine parameter along the null ray is equal to the value of the action (modulo two fixed constants: a multiplicative and an additive one).
In other words, if the physicist knows the shadows of all geodesics together with the value of this affine parameter, then she/he is able to reconstruct the ambient spacetime. This procedure provides a concrete description of the Eisenhart lift.
In a sense one might say that if the value of the action is considered as a sort of extra coordinate that one should add to the absolute space and time coordinates for the description of nonrelativistic dynamical trajectories,
then the corresponding constructed spacetime with one more dimension admits a natural description in terms of a gravitational wave.

Spacetimes with a null hypersurface-orthogonal Killing vector field have already been investigated in the literature \cite{Julia:1994bs,Minguzzi:2006gq} but, to our knowledge, no specific name has been given to this wide class of spacetimes. 
Since this is the one relevant for the ambient approach and
as a tribute for the stimulating analogy \cite{Minguzzi:2006gq} with the allegory of the cave, we will refer to such a spacetime as a ``Platonic gravitational wave''. Accordingly, its orbit space of null rays will be called ``Platonic screen''. The projection of ambient objects (such as clocks, geodesics, \textit{etc}.) on this screen will be called their ``shadows''.

\section{Schr\"{o}dinger equation from Klein-Gordon equation}
\label{Schrodinger}
As shown in section \eqref{Nonrelativistic Hamiltonian}, the Eisenhart-Lichnerowicz theorem for the classical particle acquires a simpler formulation when seen from a Hamiltonian perspective. In the present section, the theorem is extended to first-quantized equations for a scalar field, \ie the Schr\"odinger equation is derived as a null dimensional reduction of the Klein-Gordon equation. In a first step, we review the results of \cite{Duval:1984cj,Duval:2012qr} by performing the reduction for the line element \eqref{Elinel} before generalising these results to the conformally equivalent class \eqref{Llinel}.

According to the standard rules of quantification, the momenta appearing in the classical Hamiltonian formalism are essentially converted into partial derivatives and the Hamiltonian turns into an operator such that the mass-shell constraint becomes the Klein-Gordon equation. One then faces the ambiguity because of the introduction of noncommuting operators. We choose to fix the ambiguity by focusing on the conformal invariant Laplacian of Yamabe, in order to take advantage of the conformal relation between the classes of spacetimes at hand. This formalism is reminiscent of the light-cone formulation \cite{Dirac:1949cp} and can be seen as a generalization thereof to suitable curved spacetimes. 
~\\~\\Starting with the $D$-dimensional Klein-Gordon action:
\bea
S=\int{d^Dx \, \sqrt{-g}\, \Phi^*\pl\Box_Y \Phi-M^2\Phi\pr}\,,\label{KGaction}
\eea
whose equations of motion read
\bea
\Box_Y \Phi-M^2\Phi=0\,,
\eea
where $\Box_Y=\Box-\frac{D-2}{4\pl D-1\pr}R$ is the Yamabe operator, with $\Box\equiv\nabla^\mu\nabla_\mu$ the Laplace-Beltrami operator. The Yamabe operator is also known as the conformal Laplacian, because of the conformal invariance of the equation $\Box_Y \Phi=0$ (see \eg appendix D of \cite{Wald}). More precisely, if $g$ and $\g$ are conformally related via $g=\Om \g$, then the equation $\Box_Y \Phi=0$ is said to be conformally invariant with weight $-\frac{d}{4}$ (where $d=D-2$), \ie it satisfies: \bea
\Box_Y\pl\Om^{-\frac{d}{4}}\Phi\pr=\Om^{-1-\frac{d}{4}}\bar\Box_Y\Phi. \label{conformally invariance}
\eea 

We start by considering the line element \eqref{Elinel} (this class of metrics will be referred to as Bargmann-Eisenhart waves in the following sections) and perform the dimensional reduction of the action \eqref{KGaction} along the lightlike direction $\frac{\partial}{\partial u}$ by considering a specific Fourier mode in the direction $u$: $\Phi(u,t,\vec{x})=\phi(t,\vec{x})e^{imu}$. As can be easily checked, the scalar curvature and determinant of the metric \eqref{Elinel} are equal to the ones of the spatial metric $\g_{ij}$ so we have $\bar R=\bar R^{(d)}$ and $\det \g=\det \g^{(d)}$. 

The action \eqref{KGaction} then reduces to: 
\bea
S&=&\int d^Dx \, \sqrt{\g^{(d)}}\, \phi^*\Bigg[ D ^2\phi+2im\p_t\phi+\half im\p_t\pl\ln \g^{(d)}\pr\phi\\
&&\qquad\qquad\qquad\qquad-\pl M^2+2m^2\bU+\frac{d}{4\pl d+1\pr}\R^{(d)}\pr\phi\Bigg]\nn
\eea
where we introduced the covariant derivative $D _i\phi=\na^{(d)}_i-im\bA_i$. 
For cosmetic reasons, the term involving the time derivative of the determinant for the metric $\g$  can be integrated by parts to obtain:
\bea
S=\int{d^Dx \, \sqrt{\g^{(d)}}\, \pl\phi^*D ^2\phi+2m^2\rho-\pl M^2+2m^2\bU+\frac{d}{4\pl d+1\pr}\R^{(d)}\pr|\phi|^2\pr}
\eea
where $\rho$ stands for the density probability: $\rho=\frac{i}{2m}\pl\phi^*\p_t \phi-\phi\p_t\phi^*\pr$. 
The associated equations of motion then read:
\bea
\bar\Box_Y\Phi-M^2\Phi&=&e^{imu}\Big[ D ^2\phi-2m^2\bU\phi+2im\p_t\phi\nn\\
&&+\half im\p_t\pl\ln \g^{(d)}\pr\phi-\frac{d}{4\pl d+1\pr}\R^{(d)}\phi-M^2\phi\Big]=0
\eea
so that Klein-Gordon equation on the curved spacetime \eqref{Elinel} reduces to Schr\"{o}dinger equation on the curved space $\g_{ij}$ (see \textit{e.g.} \cite{Duval:1983pb}):
\bea
i\p_t\phi=\crl-\frac{1}{2m}\pl D ^2+\frac{d}{4\pl d+1\pr}\R^{(d)}\pr+m\,\bar V'-\frac{i}{4}\p_t\pl \ln\g^{(d)}\pr\crr\phi
\eea
where we defined $\bar V'=\bU+\frac{M^2}{2m^2}$. The operator $i\p_t+\frac{1}{2m}\pl D ^2+\frac{d}{4\pl d+1\pr}\R^{(d)}\pr+\frac{i}{4}\p_t\pl \ln\g^{(d)}\pr$ can be seen as a nonrelativistic equivalent of the Yamabe operator. 

We now switch to the class of metrics whose line element takes the form \eqref{Llinel} (later referred to as Platonic waves), which are conformally related to the previously studied class as we have $g=\Om\pl t,x\pr \bar g$. The choice of the Yamabe operator then turns out to be handy, thanks to the property \eqref{conformally invariance} which suggests the following ansatz: $\Phi(u,t,\vec{x})= \Omega^{-d/4}\phi(t,\vec{x})e^{imu}$ under which the action \eqref{KGaction} becomes: 
\bea
S=\int{d^Dx \, \sqrt{\g^{(d)}}}\, \phi^*\Bigg[ D ^2\phi+2im\p_t\phi+\half im\p_t\pl\ln \g^{(d)}\pr\phi\nn\\ -\pl M^2\Om+2m^2\bU+\frac{d}{4\pl d+1\pr}\R^{(d)}\pr\phi\Bigg]. 
\eea
The associated equations of motion read
\bea
&&\Box_Y\Phi-M^2\Phi=\Omega^{-1-\frac{d}{4}}e^{imu}\Big[ D ^2\phi+2im\p_t\phi\nn\\
&&\quad+\half im\p_t\pl\ln \g^{(d)}\pr\phi-\pl M^2\Omega+2m^2\bU+\frac{d}{4\pl d+1\pr}\R^{(d)}\pr\phi\Big]=0
\eea
which once again leads to Schr\"{o}dinger equation:
\bea
i\p_t\phi=\crl-\frac{1}{2m}\pl D ^2+\frac{d}{4\pl d+1\pr}\R^{(d)}\pr+m\,\bar V-\frac{i}{4}\p_t\pl \ln\g^{(d)}\pr\crr\phi
\eea
with $\bar V=\bU+\frac{M^2\Omega}{2m^2}$. 

\section{Geometric definitions of Platonic gravitational waves}
\label{Geometric definitions Plato}
 Similarly to the definition of manifolds endowed with a \textit{Riemannian} structure, \textit{i.e.} a positive-definite metric, one can define \textit{relativistic spacetimes} as smooth manifolds endowed with a \textit{Lorentzian} structure, \textit{i.e.} a metric with signature $(-,+,...,+)$\,.
 Somewhat less familiar to most physicists are the \textit{nonrelativistic spacetimes} which are
 smooth manifolds endowed with absolute clock and rulers or even absolute time and space (to be defined below).
As will be shown, gravitational waves may hide such nonrelativistic structures inside their space of rays.

The notions of a gravitational wave (defined geometrically as a spacetime with a null hypersurface-orthogonal vector field), of a Bargmann-Eisenhart gravitational wave (= with parallel wave vector field) and of a Platonic gravitational wave (= conformal to a Bargmann-Eisenhart wave and with Killing wave vector field) are introduced together with the canonical form of their metric.

\subsection{Embedding nonrelativistic structures
}\label{Brinkmann}          



The present article deals with nonrelativistic features embedded inside relativistic spacetimes. In this context, one can legitimately ask: what constitutes the most general class of relativistic spacetimes inducing a nonrelativistic structure\,? In order to address this question, one needs first to properly define nonrelativistic structures. We will at first follow the definition of \cite{Bernal:2002ph} of a Leibnizian structure, which will turn out to be too weak a requirement and next switch to the more restrictive notion of Aristotelian structure.  
~\\A \textit{Leibnizian structure} \cite{Bernal:2002ph} comprises the following three elements: a manifold $\mathcal M$
, a 1-form $\psi$ and a positive-definite metric $\gamma$ acting on the kernel of $\psi$ (Everywhere in this paper are vector fields and 1-forms assumed to be nowhere vanishing. This assumption will often be left implicit for the sake of brevity. Similarly, manifolds are taken to be smooth and connected.). We will call $\psi$ an \textit{absolute clock} and $\gamma$ a collection of \textit{rulers}. As such, it is easy to see that any relativistic spacetime induces a Leibnizian structure. Indeed, the tangent space to a $D$-dimensional relativistic spacetime is isomorphic to Minkowski spacetime and can be endowed at each point with a set of $D$ orthogonal coframes ($e_0, e_1,...,e_{D-1}$). Choosing $\psi\equiv e_0$ as an absolute clock, each point is endowed with a positive-definite metric acting on the kernel of $\psi$ engendered by the vectors dual to the forms $e_1,...,e_{D-1}$. 

As is now manifest, the above definition of a nonrelativistic structure is too weak to discriminate a subclass of relativistic spacetimes. Furthermore, it does not allow a global definition of absolute time and space since it only provides a set of local clocks and rulers. These two drawbacks of the previous definition can be circumvented by the introduction of an extra condition on the 1-form $\psi$. 
The requirement that the nonrelativistic structure allows a global notion of absolute time and space amounts to define submanifolds of $\mathcal M$ endowed with the spatial metric $\gamma$, \ie they have to admit the kernel of the 1-form $\psi$ as tangent vector space. The necessary and sufficient condition for the existence of such integral submanifolds (see \textit{e.g.} appendix B.3 of \cite{Wald}) is the Frobenius integrability condition $\psi\wedge d\psi=0$, so that the kernel of $\psi$ defines a foliation of $\mathcal M$ by a family of hypersurfaces of codimension-one called \textit{simultaneity slices}. These are the integral submanifolds endowed with the spatial metric $\gamma$. 
 Locally, $\psi=\Omega\,dt$ where $\Omega> 0$ and the function $t$ is called an \textit{absolute time}. The simultaneity slices are the hypersurfaces of fixed absolute time and are called \textit{absolute spaces}, as they are endowed with the positive-definite metric $\gamma$. 
We will call a Leibnizian structure whose absolute clock satisfies the Frobenius integrability condition an \textit{Aristotelian} structure. They were called Leibnizian structures with \textit{locally synchronizable} clock in \cite{Bernal:2002ph}.\footnote{We did not retain the reference to Leibniz because it is somewhat improper since he actually debated with Newton and strongly argued against absolute time and space. We preferred to refer to Aristotle because Aristotelian physics is pre-relativist (even in the Galilean sense) and also does not include the inertial principle. Accordingly, our definition of Aristotelian structure does not involve any notion of parallelism (contrarily to Galilean structure, \textit{c.f.} \cite{Bernal:2002ph,Duval:1984cj}). 
} 

~\\In order to determine the class of relativistic spacetimes inducing an Aristotelian structure, we seek for spacetimes admitting a hypersurface-orthogonal vector field [the dual to the absolute clock $\psi$, denoted $\xi\equiv g^{-1}\pl\psi\pr$] and restrict for simplicity our analysis to the case where $\xi$ is of definite type throughout the entire spacetime. We further restrain to cases when the transverse metric on the simultaneity slices is positive semidefinite, as seems natural in order to induce an Aristotelian structure on them (or a quotient thereof). 
As spacetimes admitting a spacelike hypersurface-orthogonal vector field necessarily induce a Lorentzian transverse metric, they do not constitute natural candidates in order to yield a positive-definite spatial metric. Therefore, we are left with the following two cases: 
\vspace{2mm}

$\bullet$ $g\pl \xi,\xi\pr<0$: Relativistic spacetimes admitting a timelike hypersurface-orthogonal vector field indeed induce an Aristotelian structure as the transverse metric to the vector field on the simultaneity slices is positive definite. This class of time-foliated spacetimes includes the Friedmann-Lema\^{i}tre-Robertson-Walker spacetimes whose cosmological time labels the different slices which are homogeneous spaces. 
A peculiarity of time-foliated spacetimes is that they possess both relativistic and nonrelativistic features, \ie the nonrelativistic spacetime merges with the relativistic spacetime, and not with a quotient thereof. This interesting class will not be considered further here, being already well studied and moreover stepping outside the scope of the present article which focuses on dimensional reduction. 
\vspace{2mm}

$\bullet$ $g\pl \xi,\xi\pr=0$: The lightlike case will constitute the main object of study of the present section and associated relativistic spacetimes will be called \textit{gravitational waves}. 
\subsubsection{Gravitational waves
}\label{GW structure}
The class of spacetimes with a null hypersurface-orthogonal vector field has the nice feature of allowing the introduction of a special chart of coordinates, the so-called \textit{Brinkmann coordinates}\footnote{The term Brinkmann coordinates seems standard for pp-waves \cite{Blau} but they were originally introduced for Bargmann-Eisenhart spacetimes \cite{Brinkmann}. Here we slightly generalize the denotation of this term. } which induce a canonical form for the metric. This is actually the chart we used in section \ref{Nonrelativistic} and which we will use extensively in the following. These spacetimes are also interesting since, as suggested by their name, they possess the minimal structure allowing a fruitful usage of wave-related features for their characterization. 

We start with some definitions: a \textit{wave vector field} is a hypersurface-orthogonal null and complete vector field, the orbits of which are called \textit{rays}.
\begin{defi}
A gravitational wave is a Lorentzian manifold possessing a wave vector field.
\end{defi}
The congruence of rays defines the gravitational wave via the standard rules of geometric optics.
For instance, a \textit{wavefront worldvolume} is a hypersurface which is orthogonal to the congruence of rays.
Wavefront worldvolumes are thus codimension-one null hypersurfaces containing a (sub)congruence of rays (because the wave vector field is orthogonal to itself), \textit{c.f.} Fig.\ref{fig2}. By definition, a gravitational wave is a spacetime foliated by the wavefront worldvolumes.

\begin{figure}[ht]
\centering
   \includegraphics[width=0.7\textwidth]{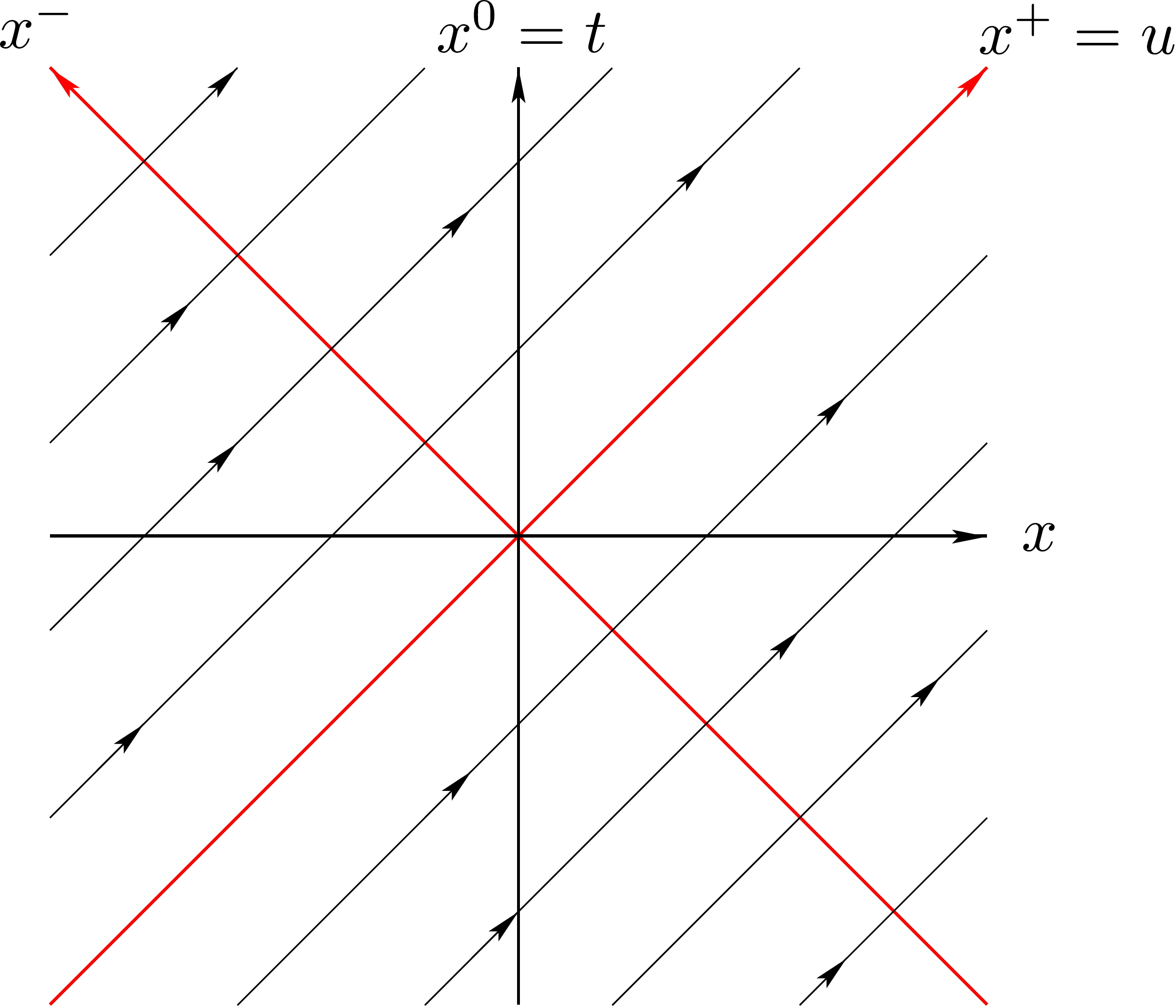}
  \caption{Two-dimensional Minkowski spacetime as a gravitational wave. The wavefront worldvolumes are the lines $x^-=\mbox{const}$.  \label{fig4}}
\end{figure}
\vspace{2mm}
\noindent\textbf{Example:}
The simplest example of a gravitational wave (according to the above definition)
is Minkowski spacetime. It can indeed be foliated by any collection of parallel null hyperplanes, interpreted as flat wavefront worldvolumes (Fig.\ref{fig4}). The corresponding congruence of rays is provided by the parallel null lines inside each leave.
\vspace{2mm}

We will denote the wave vector field by $\xi$. The differential 1-form dual to $\xi$ is referred to as the \textit{wave covector field} and written $\psi\equiv g\pl\xi\pr$, the components of which are: $\psi_\mu\equiv g_{\mu\nu}\xi^\nu=\xi_{\mu}$.
Due to the hypersurface-orthogonality condition on the wave vector field $\xi$, the wave covector field can be written locally as $\psi= {\Om}\, df$ where the primitive $f$ is called the \textit{retarded time} (or ``phase'') and we assume without loss of generality that $\Om> 0$. In components, this reads as $\xi_\mu=\Omega\, \partial_\mu f$.
As one can see, the level sets of the retarded time (\textit{i.e.} the loci $f=$ constant) are the wavefront worldvolumes.
Notice that, since the wave (co)vector field is null, ${\cal L}_\xi f=0$ (since $0=\xi_\mu \xi^\mu=\Om\, \xi^\mu\p_\mu f$).

\begin{figure}[ht]
\centering
   \includegraphics[width=0.7\textwidth]{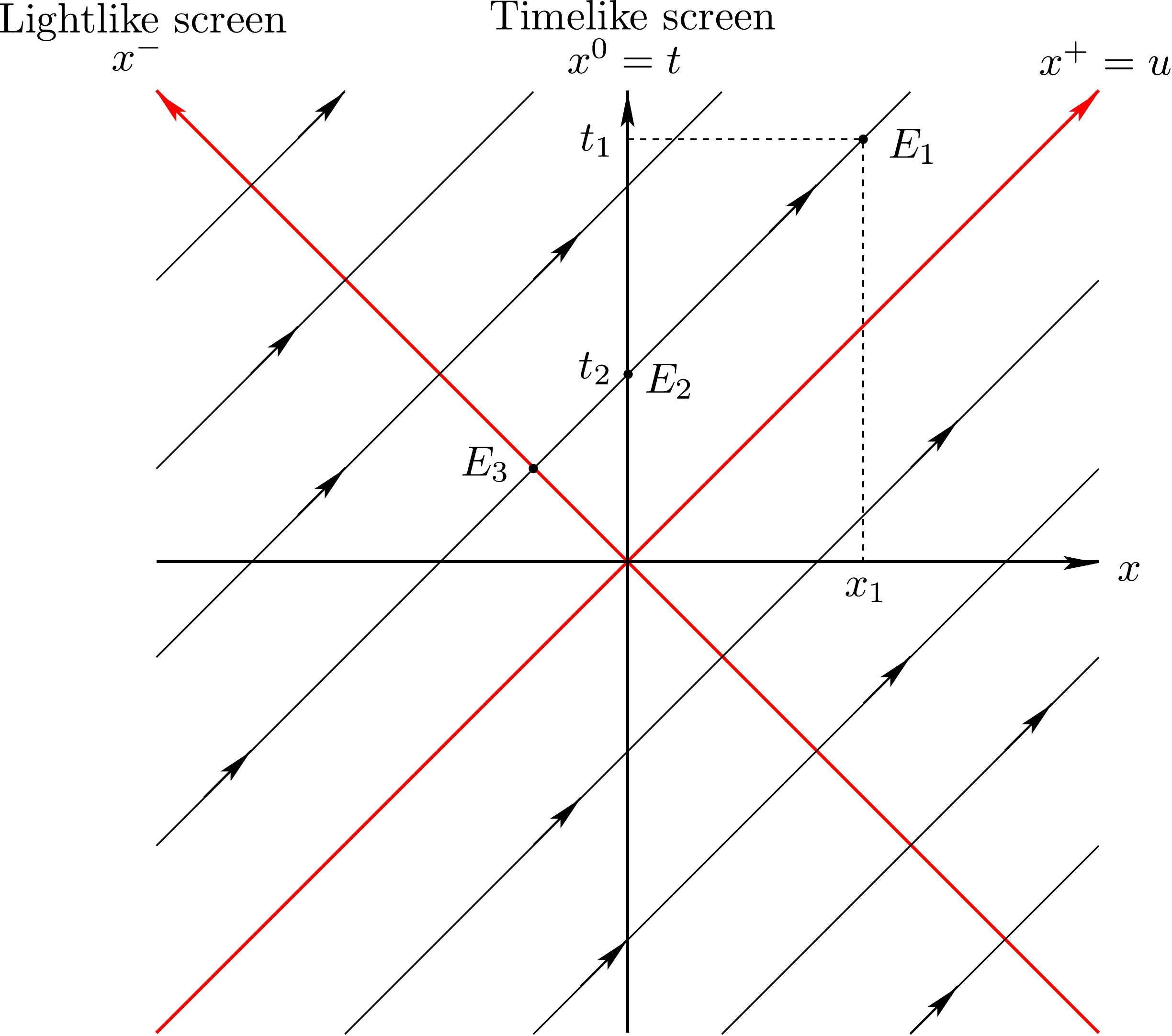}
  \caption{Several choices of screen worldvolumes are possible \textit{e.g.} the timelike screen worldvolume axis $x^0$, so that leaves of the foliation are labeled by the retarded time $t$, or the lightlike screen worldvolume axis $x^-$ which labels leaves with light-cone time $x^-$. The event $E_1$ is encoded on the timelike screen worldvolume by its position $x_1$ and the time of emission ($E_2$) of the graviton intersecting it: $t_2=t_1-x_1$. Alternatively, on the lightlike screen worldvolume, the moment of emission ($E_3$) of the graviton intersecting $E_1$ has for light-cone time $x^-=\frac{t_1-x_1}{\sqrt{2}}$. \label{fig5}}
\end{figure}

\begin{lem}
The wave covector field defines a locally synchronizable absolute clock on a gravitational wave, whose absolute time is the retarded time and whose simultaneity slices are the wavefront worldvolumes.
\end{lem}
\vspace{2mm}
\noindent\textbf{Example:} Light-cone time $x^-$ provides an absolute time on Minkowski spacetime (Fig. \ref{fig5}).
Notice in this example, that contrary to nonrelativistic spacetimes, there may exist several inequivalent ``absolute'' times (for instance $x^-$ or $x^0$ in fig.\ref{fig5}) on relativistic spacetimes that admit inequivalent wave vector fields.
\vspace{2mm}

\subsubsection{Brinkmann coordinates}\label{Plato Brinkmann}
The \textit{Brinkmann coordinates} are now introduced as follows: two among the $D=n+1$ coordinate vector fields $\frac{\partial}{\partial x^\mu}$ are specialized, let us call them $\frac{\partial}{\partial u}$ and $\frac{\partial}{\partial t}$. The first coordinate is taken to be the affine parameter $u$ along rays (so the
corresponding coordinate vector field is identified with the wave vector itself, $\frac{\partial}{\partial u}=\xi$); the second coordinate corresponds to the retarded time ($t=f$); and the remaining $d=n-1$ coordinates $x^i$ are coordinate systems on the wavefronts.\footnote{We follow the coordinate convention of \cite{Eisenhart1928} and \cite{Duval:1984cj} which differs from the standard notation in gravitational waves literature where our $(u,t)$ coordinates are usually denoted $(v,u)$ respectively.} Thus, one has $g_{u\mu}=g\pl\xi,\frac{\partial}{\partial x^\mu}\pr=\xi_\mu=\Om\, \delta^t_\mu$. From this last relation, one sees that the remaining $d=n-1$ coordinate vector fields $\frac{\partial}{\partial x^i}$ are orthogonal to the null vector field, as they should since by construction the coordinates $(u,x^i)$ must provide coordinates on the wavefront worldvolumes.
Similarly, the coordinates $(t,x^i)$ provide coordinates on the hypersurface $u=0$ that can be interpreted as a screen worldvolume corresponding to the choice of transverse vector field $\frac{\partial}{\partial t}$.
\vspace{2mm}

In a Brinkmann coordinate chart, the line element thus takes the canonical form:
$$
ds^2=g_{tt}\, dt^2\,+\,2\,\Om\, dt\, du\,+\,2\,g_{ti}\, dx^i dt\,+\,g_{ij} dx^i dx^j\,,
$$
where the metric components $g_{\mu\nu}$ are in general functions of all the coordinates.
Looking backward to section \ref{Nonrelativistic Lagrangian} or forward to section \ref{Definitions of Platonic gravitational waves}, one can introduce the (scalar) potential $\bU=-\half \Omu g_{tt}$, the Coriolis 1-form $\bA_i=\Omu g_{ti}$ and the conformally related spatial metric $\g_{ij}=\Omu g_{ij}$ and reexpress the canonical line element as:
\bea
ds^2=\Om\pl t,x\pr \crl2\,dt\pl  du+\bA_i
(u,t,x)\, dx^i-\bU(u,t,x)dt\pr \,+\,\g_{ij}(u,t,x)\, dx^i dx^j\crr\label{GW}
\eea
where, without loss of generality, $\Om$ can be taken independent of $u$, as will be shown later.
The inverse metric now reads:
\bea
g\un=\Om\un\crl\pl 2\bU+\g^{ij}\bA_i \bA_j\pr\p_u\otimes\p_u+\p_u\otimes\p_t+\p_t\otimes\p_u-\g^{ij}\bA_j\pl\p_u\otimes\p_i+\p_i\otimes\p_u\pr+\g^{ij}\p_i\otimes\p_j\crr. \label{eqmetricinv}\nn
\eea
\vspace{2mm}
\noindent\textbf{Example:}
The light-cone coordinates $x^\mu$ ($\mu=+,-,i$), where $x^\pm=(x^0\pm x^{n})/\sqrt{2}$ on the Minkowski spacetime ${\mathbb R}^{n,1}$, provide Brinkmann coordinates for the simplest instance of a gravitational wave (Fig.\ref{fig5}). 
The flat line element reads
\bea
ds^2=\eta_{\mu\nu}dx^\mu dx^\nu=-2\,dx^+dx^-+\delta_{ij}dx^idx^j\,,
\eea
so that one might identify the retarded time $t$ with $x^-$ and the affine parameter $u$ with $x^+$.
\vspace{2mm}

It will be useful for some calculations to dispose of a frame version of the Brinkmann coordinates.
A \textit{light-cone frame} is a moving (co)frame where the line element takes the form
\bea
ds^2=\eta_{ab}e^a e^b=-2\,e^+e^-+\delta_{ij}e^ie^j. \label{adapted line element}
\eea
In the Petrov-type classifications, the vectors $e^-$, $e^+$, $e^i$ are often denoted by $\ell$, $n$, $m^i$, respectively.
An \textit{adapted frame} is defined as a light-cone frame where the null frame $\ell\equiv e^-$ is taken to be
 the clock $\psi=g\pl\xi\pr$. 
 The other null (co)frame $n\equiv- e^+$ is then completely determined by the line element (\ref{adapted line element}). Often the Brinkmann coordinates will be used, so that the null coframes will read $\ell=g\pl\xi\pr=\Om\, dt$ and $n=du+\bA_i\, dx^i- \bU dt$.

\vspace{2mm}
There is no canonical prescription for the remaining ``orthonormal'' coframes $m^i\equiv e^i$ on the wavefronts, which must be such that $$\delta_{ij}e^ie^j=g_{ij}\, dx^i dx^j\,.$$
As one can see from (\ref{adapted line element}), $e^+$ and $e^-$ being null, the (co)frames $e^i$ must be spacelike in order for the spacetime metric $g_{\mu\nu}$ to have a Lorentzian signature, and so the metric $g_{ij}$ must be positive definite.\footnote{The positive-definiteness of the spatial metric $\gamma$ is also obvious from the calculation of the determinant of the ambient metric (\ref{GW}) which reads $\det\, g=-\Om^D \det\, \gamma$.} However, the type of $\frac{\partial}{\partial t}$ (\textit{i.e.} the sign of $g_{tt}$ and $\bU$) can be anything.
\vspace{2mm}

The 1-forms $\ell$ and $n$ are also useful to covariantly define the \textit{transverse metric}
\be
{}^\perp\gamma_{\mu\nu}=g_{\mu\nu}-2\, n_{(\mu}\ell_{\nu)}=g_{ij}e_\mu^ie_\nu^j
\ee
with $n^2=\ell^2=0$ and $n\cdot \ell=1$. It is easy to check that the wave vector field $\xi=\frac{\partial}{\partial u}$, as well as $\frac{\partial}{\partial t}$, belong to the kernel of ${}^\perp\gamma$. The transverse metric ${}^\perp\gamma$ is necessary in order to define the \textit{optical scalars} associated to the wave vector field $\xi$, \textit{i.e.} the expansion $\theta=\nabla^{\alpha}\xi_{\alpha}$, the shear $\sigma$ and the twist $\omega$. The transverse part of the tensor $\nabla \xi$ can indeed be decomposed into its $\mathfrak{o}(d)$-irreducible parts as ${}^\perp\gamma_{\mu}^\alpha{}^\perp\gamma_{\nu}^\beta\nabla_{\beta}\xi_{\alpha}=\frac1{d}\,\theta\,{}^\perp\gamma_{\mu\nu}+\sigma_{\mu\nu}+\omega_{\mu\nu}$ with $\sigma_{\mu\nu}=\sigma_{(\mu\nu)}$ and ${}^\perp\gamma^{\mu\nu}\sigma_{\mu\nu}=0$ and $\omega_{\mu\nu}={}^\perp\gamma_{[\mu}^\alpha{}^\perp\gamma_{\nu]}^\beta\nabla_{\beta}\xi_{\alpha}={}^\perp\gamma_\mu^\alpha{}^\perp\gamma_\nu^\beta\nabla_{[\beta}\xi_{\alpha]}$. 
The shear $\sigma$ and twist $\omega$ are the scalar fields respectively defined by
$\sigma^2=\frac12\,\sigma^{\mu\nu}\sigma_{\mu\nu}=\frac12\,\sigma^{ij}\sigma_{ij}$ and
$\om^2=\frac12\,\om^{\mu\nu}\om_{\mu\nu}=\frac12\,\om^{ij}\om_{ij}$. Since $\sigma^2$ and $\omega^2$ are sums of squares, the shear $\sigma$ and the twist $\omega$ respectively vanish if and only the tensors $\sigma_{\mu\nu}$ and $\omega_{\mu\nu}$ respectively vanish.

\vspace{2mm}
\noindent\textbf{Remark:} We stress that the ``rotational'' two-form (or ``curl'') $d\xi$ with components $\partial_{[\mu}\xi_{\nu]}=\nabla_{[\mu}\xi_{\nu]}$ and the ``rotation'' (or ``twist'') two-form $\omega$ with components $\omega_{\mu\nu}={}^\perp\gamma_\mu^\alpha{}^\perp\gamma_\nu^\beta\nabla_{[\beta}\xi_{\alpha]}$ are in general distinct tensors. Indeed, they must be distinguished for null forms, although they coincide for time (or space) like ones. In fact, from Frobenius theorem one knows that a wave vector field is automatically twistless, although it is not necessarily irrotational.
\vspace{2mm}

In the Brinkmann coordinates, the kernel of $\psi$ at each point of the simultaneity slices is the $n$-dimensional vector space composed of tangent vectors $X$ satisfying $g\pl X,\frac{\p}{\p t}\pr=0$. Therefore, for $X,Y$ belonging to the kernel of $\psi$, the action of $g$ writes $$g\pl X,Y\pr=g_{ij}X^i Y^j=\gamma\pl X,Y\pr=
{}^\perp\gamma\pl X,Y\pr$$ so the induced (or transverse) metric ${}^\perp\gamma$ on the simultaneity slices is of rank $d=n-1$ and its action reduces to the one of the positive-definite $d$-dimensional spatial metric $\gamma$. 
The wavefront worldvolumes are then endowed with a positive semidefinite metric ${}^\perp\gamma$ and then, as such, cannot be given the interpretation of absolute spaces. In order to obtain a nondegenerate metric, one can quotient the wavefront worldvolume by the null direction. However, this procedure is only well-defined if the rays are orbits of an isometry. As we will argue, this further requirement is necessary in order for a gravitational wave to induce an Aristotelian structure. The next subsection is devoted to a description of this quotient manifold. 

\subsubsection{Platonic screens}\label{Platonic screens}
The previous ``gravitational wave'' terminology is further justified when one considers the following lemma:
\begin{lem}
Any wave vector field is geodesic.
\end{lem}
\noindent Consequently, rays are null geodesics and can thus be interpreted as graviton worldlines.

\proof{Using the hypersurface-orthogonality of the vector field and Frobenius theorem, we see that the 1-form $\psi\equiv g\pl\xi\pr$ satisfies $d\psi=\alpha\wedge\,\psi$ for some 1-form $\alpha$. Expressing the left-side in terms of covariant derivatives and contracting with $\xi$, one obtains $\nabla_\xi \xi-\half \nabla\pl\xi^2\pr=\pl\alpha\,\cdot\, \xi\pr\xi- \pl\xi^2\pr\alpha$, which, for a null vector field ($\xi^2=0$), is equivalent to the geodesic condition\footnote{A comment on the terminology is in order. In this work, the term \textit{geodesic} will be used to designate not-necessarily affinely-parametrized geodesic vector fields (\textit{i.e.}
satisfying $\nabla_\xi \xi=\kappa\, \xi$ with $\kappa$ a function of coordinates) and prefer the term \textit{affine geodesic} for affinely-parametrized vector fields (satisfying $\nabla_\xi \xi=0$). }. }


Without loss of generality, the wave vector field $\xi$ will be taken to be 
affine geodesic
from now on. The equation $\nabla_\xi \xi=0$ implies that ${\cal L}_\xi\Om=0$ (as can be obtained from the local expression of the curl of the wave covector,
$\partial_{[\mu}\xi_{\nu]}=\partial_{[\mu}\Om\, \partial_{\nu]}f$,
expressed in terms of covariant derivatives and contracted with the vector field $\xi^\mu$). As mentioned above, the factor $\Omega$ is thus independent of the affine parameter $u$ along rays. 

This property is important in order for $\Omega$ to acquire the interpretation of a time unit on the quotient manifold defined as follows: 
\begin{defi}
The Platonic screen is the orbit space of rays for a gravitational wave, \textit{i.e.} the points of the Platonic screen are identified with the rays of the gravitational wave.
\end{defi}
There is no canonical realization of the Platonic screen as a submanifold of the gravitational wave since various slicing{s} are perfectly legitimate. However, any such slicing corresponds to a specific choice of representative in each orbit.
These subtleties
justify the rather abstract but geometric definition of the Platonic screen. A \textit{screen worldvolume} is a submanifold of a gravitational wave providing a complete set of representatives of the Platonic screen. In other words, the points of a screen worldvolume are representatives of equivalence classes constituted by the rays (Table \ref{table2}). In some sense, any screen worldvolume can be seen as a concrete realization of the abstract Platonic screen (Fig.\ref{fig1}). 
\begin{table}
\begin{center}
\begin{tabular}{
|c|c|c|c|}
 \hline
 & Spacetime & Coordinates & Structure \\
\hline\hline
 & & & \\
Manifold & Ambient spacetime & $(u,t,x^i)$ & Lorentzian \\
 & & & \\
\hline
Quotient manifold & Platonic screen & &  \\
 & & $(t,x^i)$ & Aristotelian \\
Submanifold & Screen worldvolume, \eg $u=0$ & & \\
\hline
\end{tabular}
\end{center}
\caption{Summary of the spacetimes in the ambient approach\label{table2}}
\end{table}
\begin{lem}The Platonic screen is 
endowed with a locally synchronizable absolute clock. The absolute time on any screen worldvolume is induced from the retarded time of the gravitational wave.
\end{lem}
\proof{The absolute clock locally reads $\psi=\Omega(t,x)dt$ which is well-defined on the Platonic screen, in the sense that it does not depend on the choice of screen worldvolume since the time unit $\Omega$ does not depend on the affine parameter $u$, as was shown previously.}

Similarly to the abstract definition of the Platonic screen, one defines a \textit{wavefront}
as the orbit space of rays of a wavefront worldvolume.
Again it can also be defined more concretely by the intersection between a wavefront worldvolume and a screen worldvolume, intersection which will be called a  screen (Fig.\ref{fig2}).
In other words, a \textit{screen} is a submanifold of a wavefront worldvolume providing a complete set of representatives of the wavefront (Table \ref{table3}). 
\begin{table}
\begin{center}
\begin{tabular}{
|c|c|c|c|}
 \hline
 & Leave & Coordinates & Signature \\
\hline\hline
 & & & \\
Manifold & Wavefront worlvolume $t=$const & $(u,x^i)$ & Null \\
 & & & \\
\hline
Quotient manifold & Wavefront & &  \\
 & & $(x^i)$ & Riemannian \\
Submanifold & Screen $t=$const and \eg $u=0$ & & \\
\hline
\end{tabular}
\end{center}
\caption{Summary of the leaves in the ambient approach\label{table3}}
\end{table}
\begin{figure}[ht]
\centering
\includegraphics[width=0.7\textwidth]{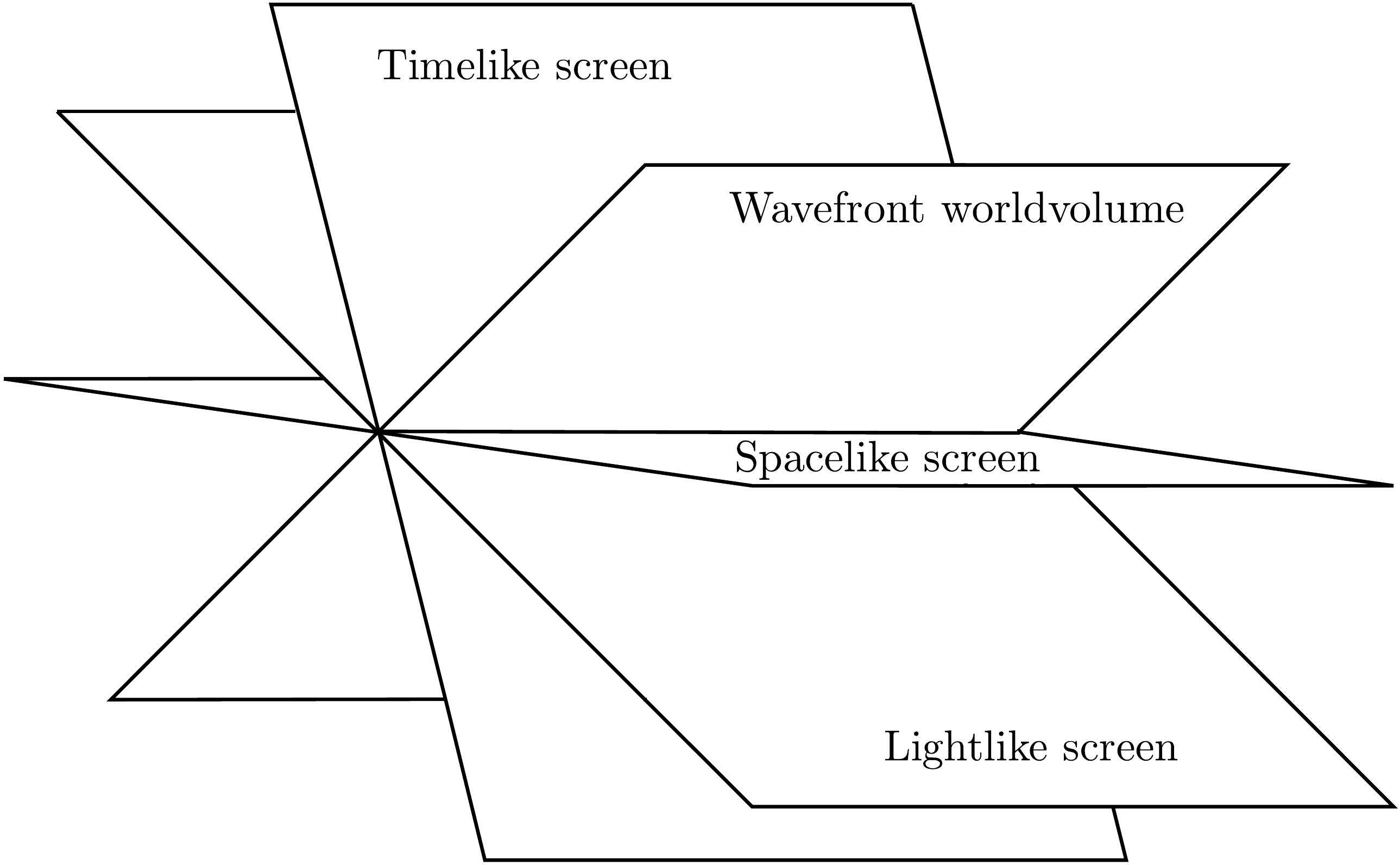}
\caption{\label{Different gender} Examples of screens of different types\label{fig6}}
\end{figure}
 A smooth choice of representatives for the complete set of wavefront worldvolumes defines a screen worldvolume. As a side remark, let us notice that the screen worldvolumes can be of any type. 
When the context makes it clear, screen worldvolumes will sometimes be improperly referred to as ``screen'' for the sake of concision (as in Fig.\ref{fig6}). For instance, the Platonic screen actually corresponds to an infinite collection of equivalent screen worldvolumes, only differing by the choice of representatives along the rays. 

\subsection{Definitions of Platonic gravitational waves}
\label{Definitions of Platonic gravitational waves}

\subsubsection{Bargmann-Eisenhart waves}
\label{Bargmann-Eisenhart spacetimes}

Of high interest is the class of gravitational waves with parallel rays.
Precisely this class of metrics was considered by Eisenhart \cite{Eisenhart1928} in his description of dynamical trajectories as geodesic motions, so these spacetimes are sometimes called ``Eisenhart spacetimes'' by mathematicians (see \textit{e.g.} \cite{Lichnerowicz,Szydlowski}). However, the bridge between nonrelativistic physics and general relativity was rediscovered independently much later and considerably generalized in
 \cite{Duval:1984cj,Duval:1990hj} where such spacetimes were called ``Bargmann spacetimes'' in order to stress the natural appearance of the Bargmann group \cite{Bargmann:1954gh} as the structure group in this setting.
Therefore, as a tribute to both prestigious men, we will refer to these spacetimes as ``Bargmann-Eisenhart''\footnote{As a side historical remark, these spacetimes were considered by \cite{Brinkmann} so they are also sometimes called ``Brinkmann'' spacetimes \cite{GarciaParrado:2005yz}.}. 

\begin{defi}
A Bargmann-Eisenhart wave is a Lorentzian manifold with a parallel null vector field.
\end{defi}
In this subsection, the ambient metric will be denoted $\g$ in agreement with the line element \eqref{Elinel}. As suggested by our choice of terminology, these spacetimes are indeed gravitational waves. This can easily be seen as follows. The null vector field, being parallel, is necessarily curl-free and then the associated 1-form $\g\pl \xi\pr$ is closed; thus, $\xi$ is (trivially) hypersurface-orthogonal. Therefore any parallel null vector field is a wave vector field and the wave covector field is closed.

\vspace{2mm}
\noindent\textbf{Example:} It looks somehow natural to look for examples among maximally symmetric spacetimes, but this is deceptive because Minkowski spacetime is the only maximally-symmetric Bargmann-Eisenhart wave.
Indeed, spacetimes with a nonvanishing constant curvature do not admit parallel vector fields. 
\vspace{2mm}

Since Bargmann-Eisenhart waves are gravitational waves, one can
use the Brinkmann coordinates in order to bring their line element in its canonical form\footnote{We closely follow the discussion in the section 2.2 from the lecture notes \cite{Blau}.}. 
Following the prescription sketched in section \ref{Brinkmann}, one identifies $\frac{\p}{\p u}$ with the null vector field $\xi$. Being parallel, $\xi$ is also Killing and one has $\mathcal{L}_\xi \g=0$, that is, all components of the metric $\g$ are independent of the coordinate $u$. Furthermore, locally $\g\pl\xi\pr=df$ (since the wave covector field is closed) and, identifying the phase $f$ with the coordinate $t$, one obtains $\Om=1$.
The line element of a Bargmann-Eisenhart wave then takes the canonical form:
\bea
d\bar s^2&=&\g_{tt}(t,x)\, dt^2\,+\,2\,dtdu+2\g_{ti}
(t,x)\, dx^i dt\,+\,\g_{ij}(t,x)\, dx^i dx^j\nn\\
&=&2\,dt\pl du+\bA_i
(t,x)\, dx^i-\bU(t,x)dt\pr \,+\,\g_{ij}(t,x)\, dx^i dx^j\label{BEcanonical}
\eea
where in the second equation one introduced the \textit{scalar potential} $\bU=-\half \g_{tt}$, the \textit{Coriolis 1-form} $\bA_i=\g_{ti}$ (also called \textit{vector potential}) and the spatial metric $\g_{ij}$. This choice of terminology essentially follows the common usage in the Bargmann framework \cite{Duval:2008jg}.
We will also refer to the coordinate $t$, that is the primitive of the parallel null vector field as the absolute time (called ``Galilean'' time in \cite{Duval:1984cj,Duval:1990hj}), because of its nonrelativistic interpretation in the Aristotelian structure. 
On flat spacetime $\pl \bU=\bA_i=0, \, \g_{ij}=\delta_{ij}\pr$, the absolute time is identified with the light-cone time which is a null coordinate but one should keep in mind that, in general, the coordinate vector field $\partial/\partial t$ corresponding to the absolute time itself can be of any type. The arbitrariness of the signature of the screen worlvolume $u=0$ befalls to the arbitrariness of the type of $\partial/\partial t$, as can be seen from
the screen worldvolume line element \eqref{rhsmet}. It is quite remarkable that the ambient spacetime, obtained from a nonrelativistic spacetime by adding an extra coordinate $u$ and endowed with line element \eqref{BEcanonical}, has always a Lorentzian signature, despite the arbitrariness on the type of the direction $t$.

\vspace{2mm}
~\\The canonical form of the line element is preserved by local Abelian 
gauge transformations along the null fiber ($u\mapsto u-\Lambda(t,x)$, $\bU\mapsto \bU-\partial_t\Lambda$, $\bA_i\mapsto \bA_i+\partial_i\Lambda$) and by coordinate transformations of the last $d=n-1$ coordinates ($x^i\mapsto x^{'i}(t, x)$, $\bU\mapsto \bU-\half \bA_i\frac{\p x^i}{\p t'}- \bg_{ij}\frac{\p x^i}{\p t'}\frac{\p x^j}{\p t'}$, $\bA_i\mapsto  \bA_j \frac{\p x^j}{\p x^{i'}}+\g_{kl}\frac{\p x^k}{\p t'}\frac{\p x^l}{\p x^{i'}}$,
$\g_{ij}\mapsto \g_{kl}\frac{\p x^k}{\p x^{i'}}\frac{\p x^l}{\p x^{j'}}$).
\begin{figure}[ht]
\centering
\includegraphics[width=1\textwidth]{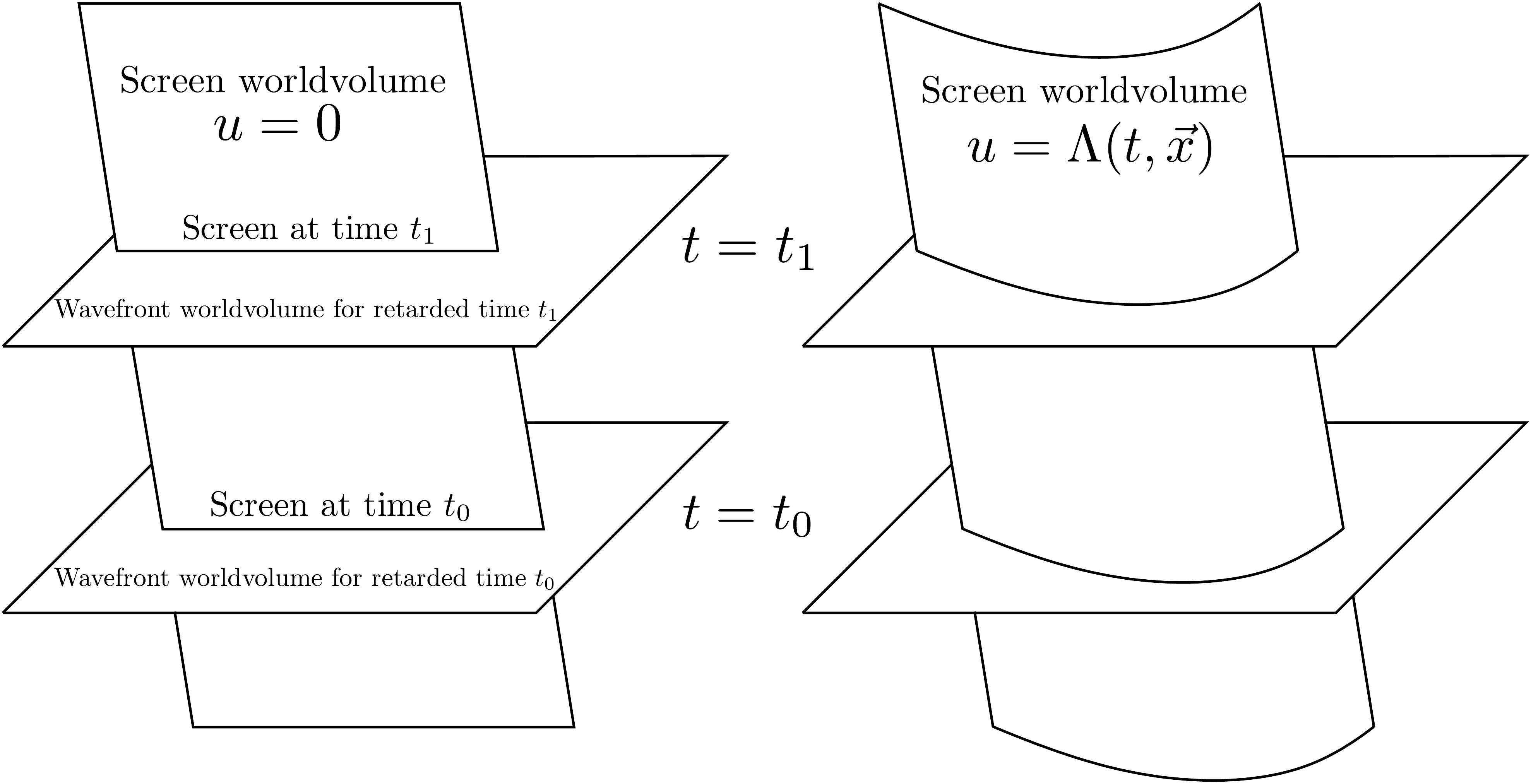}
\caption{\label{Screen worldvolume} Gauge transformation of $u$ relate different choices of screen worldvolume. \label{fig7}}
\end{figure}
While the second transformations correspond to coordinate transformations on the wavefronts, the first transformations correspond to the arbitrariness in the choice of the origin for the affine parameter along the rays. Physically, these transformations correspond to different choices of the screen worldvolume, from say the hypersurface $u=0$ to the hypersurface $u'=u-\Lambda(t,x)=0$ (\textit{c.f.} Fig. \eqref{fig7}).
Let us point out that the previous transformations also have a nonrelativistic interpretation. For instance, the Abelian gauge transformations correspond
to equivalence relation between Lagrangians differing by a total derivative, as mentioned at the end of section \ref{Nonrelativistic Lagrangian}.
Moreover, the coordinate transformations on the wavefronts correspond to the
reparametrization \eqref{repar} of holonomic coordinates. 

Furthermore, locally, it can be shown (see \textit{e.g} the section 10.1 of \cite{Ortin}) that one of the potentials, either the scalar or the vector one, can be put to zero by a suitable coordinate transformation:
\bea
u&=&u'+f\pl t',x'\pr\,,\nn\\
x^{i}&=&x^{i}\pl t',x'\pr\,,\nn
 \eea
corresponding to the following redefinitions
\bea
\bU'&=& \bU-\frac{\p f}{\p t'}-\half \bA_i\frac{\p x^i}{\p t'}- \bg_{ij}\frac{\p x^i}{\p t'}\frac{\p x^j}{\p t'}\,,\\
\bA_i'&=&  \frac{\p f}{\p x^{i'}}+\bA_j \frac{\p x^j}{\p x^{i'}}+\g_{kl}\frac{\p x^k}{\p t'}\frac{\p x^l}{\p x^{i'}}\,,\\
\g_{ij}'&=&\g_{kl}\frac{\p x^k}{\p x^{i'}}\frac{\p x^l}{\p x^{j'}}.
 \eea
It seems plausible that in fact both potentials can be set to zero, $\bU'=\bA_i'=0$, as is natural since we have as many arbitrary functions ($f$ and $x^{i}$) as potentials ($\bU$ and $\bA_i$) at our disposal; however, we are not aware of any rigorous proof of this expectation.

The curvature two-form $\bar F_{ij}=\partial_{[i}\bar A_{j]}$ of the Coriolis 1-form is called the \textit{Coriolis two-form}.
A Bargmann-Eisenhart wave whose Coriolis 1-form vanishes 
will be called \textit{Coriolis-free}.

Let us turn back now to nonrelativistic structures and see in which sense the Platonic screen of a Bargmann-Eisenhart wave is a nonrelativistic spacetime with an Aristotelian structure. 
As mentioned earlier, the wavefront worldvolumes of a gravitational wave are not absolute spaces since they are null hypersurfaces (the induced metric ${}^\perp\gamma$ is degenerate on the wavefront worldvolumes) and so although gravitational waves may induce a (locally synchronizable) absolute clock on the wavefront worldvolumes, they lack the necessary structure to define absolute spaces. However, the wavefronts of Bargmann-Eisenhart waves are Riemannian manifolds (so the Platonic screen possesses an absolute space).
In order to see why, notice that since the coordinate vector fields $\frac{\partial}{\partial x^i}$ are orthogonal to $\frac{\partial}{\partial u}$,
the (induced) metric on a wavefront is well-defined on the orbits. The tangent vectors to the wavefront
are equivalence classes $[v]$ of vectors $v\sim v+\alpha\,\xi$ ($\alpha\in\mathbb R$) and the Killing property of $\xi$ ensures that the induced metric is constant along rays.
Very concretely, the components of the positive-definite metric on the wavefronts read $\g_{ij}(t,x)$ in the Brinkmann coordinates.

We now reformulate the very beginning of the section 2 in \cite{Duval:1984cj} with our own terminology:
\begin{lem}
The Platonic screen of a Bargmann-Eisenhart wave is a nonrelativistic spacetime, where the Aristotelian structure is induced from the ambient metric.
\end{lem}

Let us now focus on a subclass of Bargmann-Eisenhart spacetimes introduced by Brinkmann \cite{Brinkmann} and vividly studied since: the so-called pp-waves.
A gravitational wave is \textit{plane-fronted} if the wavefronts define an absolute space which is flat.
Similarly, a \textit{Lobachevsky-plane-fronted wave} is a gravitational wave
where wavefronts are Lobachevsky planes \cite{Siklos} (or hyperbolic spaces in higher dimensions).
\begin{defi}\label{ppdef}
The term pp-wave stands for plane-fronted wave with parallel rays (or propagation) and designates a spacetime admitting a parallel null vector field such that the wavefronts are flat.
\end{defi}
A widespread -- though slightly misleading -- terminology defines
pp-waves as what we called Bargmann-Eisenhart waves (see \textit{e.g.} section 10.1 of \cite{Ortin}).
The reason behind this choice of terminology is the fact it implicitly assumes that only solutions of vacuum Einstein equations are considered.
Indeed, Bargmann-Eisenhart waves which are Ricci-flat
are plane-fronted in ``low'' dimensions $D\leqslant 5$, since they have a Ricci-flat spatial metric which, for $d\leqslant 3$,
is consequently flat. Moreover, Ricci-flat 
pp-waves are (essentially, \textit{c.f.} discussion below) Coriolis-free. Presumably for this reason, pp-waves in the sense of the literal definition \ref{ppdef} were called ``gyratons'' in \cite{Frolov:2005in}. As suggested by this terminology, (nonvanishing) Coriolis covector field somewhat encodes gyroscopic effects.
\vspace{2mm}

In a Brinkmann coordinate system with Cartesian coordinates on the wavefront, the line element of a pp-wave takes the canonical form:
\bea
ds^2&=&2\,dt\pl du+\bA_i
(t,x)\, dx^i-\bU(t,x)dt\pr \,+\,a^{-2}(t)\delta_{ij}\, dx^i dx^j
\eea
since here each wavefront is a flat Riemannian manifold (= Euclidean space) by assumption,
\textit{i.e.} the metric $\bar g_{ij}(t,\vec x)$ is flat for fixed absolute time $t$.
However, the coordinate transformation $\vec x'=a^{-1}\vec x$ preserving the canonical form of the metric allows us to assume without loss of generality that the canonical form of the pp-wave metric is
\bea
ds^2&=&2\,dt\pl du'+\bA_i'(t,\vec x')\, dx'{}^i-\bU'(t,\vec x')dt\pr \,+\,\delta_{ij}\, dx'{}^i dx'{}^j. 
\eea

\vspace{2mm}

We now establish that Einstein pp-waves are (under topological assumptions on the wavefront) Coriolis-free. We start by noting that the Ricci scalar of a Bargmann-Eisenhart wave is equal to the one of the wavefront, and therefore vanishes for pp-waves. Einstein pp-waves are then necessarily Ricci-flat. We now establish the following lemma: 
\begin{lem}
When the first Betti number of its wavefront is zero, a gravitational wave with zero Coriolis force ($\bar F=0$) is Coriolis-free.  
\end{lem}	

\proof{If the Coriolis force vanishes, then the Coriolis 1-form is closed ($\bar F=\bar d\bA=0$) with respect to the spatial de Rham differential $\bar d:=dx^i\partial_i$. Furthermore, if the first Betti number of the wavefront is zero, then
the Coriolis 1-form is exact ($\bA=\bar df$) and then can be gauged away via a local abelian transformation along the fiber.}
Making use of this lemma, we establish the following proposition: 
\begin{prop} When the first and second Betti numbers of its wavefront are zero, an Einstein pp-wave is Coriolis-free. 
\end{prop}
\proof{The spatial 2-form $\bar F$ on the wavefront is exact by definition ($\bar F=\bar d\bA$), thus it is closed ($\bar d\bar F=0$).
 The Ricci equation $R_{-i}=0$ implies that $\bar F$ is also coclosed ($\star\, \bar d\star \bar F=0$).  When the second Betti number of the wavefront is zero,
there are no harmonic 2-forms on it. Therefore, the Coriolis curvature is vanishing. When the first Betti number of the wavefront is zero,
this implies the Coriolis-freeness.}
Coriolis-free pp-waves then occupy a distinguished place among Bargmann-Eisenhart spacetimes. In fact, we can show that Coriolis-free pp-waves are Kerr-Schild spacetimes, a class of metrics we now briefly review.  
We will refer to a (generalized) Kerr-Schild spacetime as a manifold endowed with a metric of the following form: $g_{\mu\nu}={\mathfrak g}_{\mu\nu}-2\,{\mathfrak U}\,\xi_\mu\xi_{\nu}$, with $\xi$ a null vector field and ${\mathfrak g}_{\mu\nu}$ a constant curvature background. In flat four dimensional spacetime, this class was studied in \cite{KS} by Kerr and Schild, and was generalized  to  higher dimensions in \cite{Ortaggio:2008iq} and to (A)dS backgrounds in \cite{Malek:2010mh} where the following properties have been shown in full generality:

\vspace{2mm}

$\bullet$ The inverse metric takes the (exact) form: $g^{\mu\nu}={\mathfrak g}^{\mu\nu}+2\,{\mathfrak U}\,\xi^\mu\xi^{\nu}$ (and $|g|=1$ for flat background).

\vspace{2mm}

$\bullet$ The vector field $\xi$ is null or geodesic (or even affine geodesic) equivalenty with respect to $g$ or $\mathfrak g$.

\vspace{2mm}

$\bullet$ The expansion, shear and twist are the same with respect to $g$ or $\mathfrak g$.

\vspace{2mm}

$\bullet$ If the potential ${\mathfrak U}\,$ of a Kerr-Schild spacetime is constant along the affine geodesic null vector field, then the latter is Killing (or even parallel) equivalently with respect to $g$ or $\mathfrak g$.
\vspace{2mm}

From the above canonical form, we see that Coriolis-free pp-waves $\pl \bA_i=0\pr$ are Kerr-Schild spacetimes with Minkowski background metric: $ g_{\mu \nu}=\eta_{\mu \nu}-2{\mathfrak U}\xi_\mu\xi_\nu$. In Brinkmann coordinates the Minkowski metric reads $ {ds}^2=2\,dtdu\,+\,d\vec x^2$, while the Kerr-Schild potential is identified with the pp-wave potential ${\mathfrak U}\equiv \bU$ and $\xi=\frac{\p}{\p u}$ is the null parallel vector field.

~\\A well-known property of the Kerr-Schild spacetimes is the fact that their fully nonlinear Einstein equations reduce to their linearization around the background metric $\mathfrak g$ \textit{i.e.} Kerr-Schild spacetimes linearize the Einstein tensor. This feature greatly simplifies the equations of motion. Accordingly, for Coriolis-free pp-waves the vacuum Einstein equations reduce to the linear Laplace equation for the potential $\bU$ and, as such, Coriolis-free pp-waves traveling along the same direction are seen to obey to a superposition principle.  

\vspace{2mm}

\noindent\textbf{Examples of Coriolis-free pp-waves:}

\vspace{2mm}
\noindent $\bullet$ An \textit{exact plane wave} is a Coriolis-free pp-wave whose scalar potential is a quadratic form in the Cartesian coordinates $x^i$.
The line element of an exact plane wave then takes the form:
\bea
ds^2&=&2\,dt\pl du-M_{ij}\pl t\pr x^ix^jdt\pr\,+\,\delta_{ij}\, dx{}^i dx{}^j
\eea
with $M_{ij}(t)$ an arbitrary symmetric $d\times d$ matrix. 
\vspace{2mm}

\noindent A \textit{homogeneous plane wave} is an exact plane wave whose quadratic form is independent of the absolute time. 
A homogeneous plane wave whose matrix $M$ is proportional to the identity is a \textit{homogeneous pp-wave (Hpp-wave)}. Hpp-waves have been studied in the null dimensional reduction framework in \cite{Gibbons:2003rv} where they were shown to induce nonrelativistic spacetimes with cosmological constant (Newton-Hooke spacetimes), whose symmetry group is that of the harmonic oscillator. 

\noindent Exact plane waves are well known to enjoy the following two properties:
\begin{itemize}
\item An exact plane wave is conformally flat if and only if it is a Hpp-wave. 
Indeed, the only nonvanishing component of the Weyl tensor of a Coriolis-free pp-wave reads, in Brinkmann coordinates $C_{-i-j}=\p_i\p_j \bU-\frac{1}{d}\delta_{ij}\p_k\p^k \bU$. Substituting $\bU=M_{ij}x^ix^j$, one obtains the following condition for the matrix $M$ in order for the exact plane wave to be conformally-flat: $M_{ ij}=\frac{1}{d}\delta_{ij}M_k^k$ and $M$ is therefore proportional to the identity: $M_{ij}=\alpha\pl t\pr \delta_{ij}$ with $\alpha$ an arbitrary function of $t$. The graph of the potential of a conformally-flat plane gravitational wave is therefore a paraboloid of revolution.

\item The most important property of exact plane waves, that gave their name, is that they are Einstein manifolds if and only if their quadratic form is traceless.

As we noted, demanding that a pp-wave is an Einstein manifold is then equivalent for it to be Ricci-flat. The only nonvanishing component of the Ricci tensor of a Coriolis-free pp-wave in Brinkmann coordinates reads: $R_{--}=\p_k\p^k \bU$. Substituting $\bU=M_{ij}x^ix^j$, we see that the Ricci-flat condition is satisfied if and only if $M$ is traceless. A traceless symmetric $d\times d$ matrix indeed parametrizes the (transverse) polarization states of an on-shell \textit{linearized} gravitational wave.
We saw that this property remains manifest at nonlinear level for the Ricci-flat plane gravitational wave. 
\end{itemize}
\vspace{2mm}

\subsubsection{Platonic waves as conformal Bargmann-Eisenhart waves with preserved null Killing vector}
\label{Platonic}

The following definition of a Platonic wave is motivated by the most general form \eqref{conformalBargmann} of the line element for which the null dimensional reduction works. Its goal is to explain the geometric origin of the line element considered by Lichnerowicz  \cite{Lichnerowicz} and their relation with Bargmann-Eisenhart waves.
Later on, an equivalent definition will be provided that displays an explanation for the fact that their Platonic screen carries a structure of nonrelativistic spacetime.

\begin{defi}
Platonic waves are Lorentzian manifolds with a null Killing vector field such that the latter becomes parallel with respect to a conformally equivalent metric.
\end{defi}
As suggested by our choice of terminology, they are indeed gravitational waves: their null Killing vector field
is a wave vector field, as explained below.
The definition should be understood in more concrete terms as follows: let $\xi$ denote the null Killing vector field
with respect to the metric $g$, \textit{i.e.} ${\cal L}_\xi g=0$. The further hypothesis is that there exists a conformally related metric $\g$, that is to say $g=\Om \g$, such that $\bar\nabla \xi=0$, where $\bar\nabla$ is the covariant derivative with respect to $\g$.

As is clear from the previous definition, a Platonic wave is conformally related to a Bargmann-Eisenhart wave, both sharing the same null Killing vector field (${\cal L}_\xi g=0={\cal L}_\xi \bar g$) since a parallel vector field is automatically Killing. Hence a number of properties of Platonic waves will be easily derived from those of Bargmann-Eisenhart manifolds. Obviously, any Bargmann-Eisenhart wave is trivially a Platonic wave.

\vspace{2mm}
\noindent\textbf{Examples:} It is natural to look again for examples among maximally symmetric spacetimes.
Minkowski spacetime is of course a Platonic wave since it is even a Bargmann-Eisenhart wave.
Surprisingly enough, de Sitter spacetime is \textit{not} a Platonic wave since it does not admit a Killing vector field which is globally null (not only at the Killing horizon).
So the simplest example of a proper Platonic wave (``proper'' in the sense that it is \textit{not} a Bargmann-Eisenhart wave) is anti de Sitter spacetime.
\vspace{2mm}

Before writing the canonical form of the Platonic metric in Brinkmann coordinates, we first check that Platonic waves are gravitational waves.
The proof rests on the one for Bargmann-Eisenhart waves, where we established that the 1-form dual to the null vector field $\xi$ by the Bargmann-Eisenhart metric $\g$ is locally exact: $\g\pl \xi\pr=df$. Therefore, the 1-form obtained via the conformally related metric $g=\Om\, \g$ writes locally $g\pl \xi\pr=\Om\, df$ and $\xi$ indeed is hypersurface-orthogonal.

For later purposes, let us establish the following facts:
\begin{lem}
Two conformally equivalent spacetimes possess the same Killing vector field if and only if the conformal factor is constant along this vector field.
\end{lem}
\proof{The proof is quite straightforward: one makes use of the vanishing of the Lie derivative of the metric along a Killing vector field and of the Leibniz rule. This implies that the conformal factor $\Omega$ satisfies ${\cal L}_\xi \Omega=0$ (similarly to ${\cal L}_\xi f=0$ for any gravitational wave).}

\begin{prop}
For any Platonic wave:

$\bullet$ The conformal factor that relates it to a Bargmann-Eisenhart spacetime is constant along the null Killing vector field.

$\bullet$ The null Killing vector field is hypersurface-orthogonal and its integrating factor is equal to the conformal factor. So both the primitive and the integrating factor are constant along the null Killing vector field.

\end{prop}

\proof{A vector field is parallel if and only if it is Killing (so the lemma implies the first point) and {curl-free} (which shows the second point, since hypersurface-orthogonal is equivalent to conformally-{curl-free}).}

This justifies the use of Brinkmann coordinates and explains the form of canonical line element of Platonic waves:
\bea
ds^2&=&\,g_{tt}(t,x)\, dt^2\,+\,2\,\Om(t,x)\,dtdu\,+\,2g_{ti}(t,x)\, dx^i dt\,+\, g_{ij}(t,x)\, dx^i dx^j\nonumber\\
&=&\Om\pl t,x\pr \crl2\,dt\pl  du+\bA_i
(t,x)\, dx^i-\bU(t,x)dt\pr \,+\,\g_{ij}(t,x)\, dx^i dx^j\crr.\label{Plato} 
\eea
The second equation emphasizes the interpretation of Platonic waves as conformal Bargmann-Eisenhart waves.
In order to obtain this canonical form, one can also repeat the argument used in section \ref{Platonic screens} and use the independence of the Platonic metric from the coordinate $u$ since it corresponds to a Killing direction.

\vspace{2mm}
\noindent\textbf{Remark:} A spacetime conformally equivalent to a Bargmann-Eisenhart wave via a conformal factor that only depends on the absolute time is itself a Bargmann-Eisenhart wave admitting the same null parallel vector.
As shown in \cite{Duval:1990hj}, the converse is also true: two Bargmann-Eisenhart waves are conformally equivalent if and only if the conformal factor that relates them only depends on the absolute time.
The metric of such a spacetime (with conformal factor $\Omega(t)$) can always be put in the canonical form \eqref{BEcanonical} via a redefinition of $t$ of the form $t\mapsto t'=\int^t \Omega(\tau)d\tau$, $dt'=\Omega(t)dt$:
 \bea
ds^2&=&\Om(t)\,\left[\,\bU(t,x)\, dt^2\,+\,2\,dtdu\,+\,2\bA_i(t,x)\, dx^i dt\,+\,\g_{ij}(t,x)\, dx^i dx^j\,\right]\nn\\
&=&2\,dt'\pl du+\bA'_i
(t',x)\, dx^i-\bU'(t',x)dt'\pr \,+\,\bg'_{ij}(t',x)\, dx^i dx^j\nn
\eea
with $\bU'(t',x)=\Omu(t)\bU(t,x)$, $\bA'_i(t',x)=\bA_i(t,x)$ and $\g'_{ij}(t',x)=\Om(t)\g_{ij}(t,x)$.

\subsubsection{Platonic gravitational waves as Julia-Nicolai spacetimes}
 \label{Julia-Nicolai}

We now show the equivalence between the Platonic waves introduced in the previous subsection and the class of spacetimes studied by Julia and Nicolai in \cite{Julia:1994bs}.
\\To do so, we proceed in two steps: firstly, by reviewing the equivalence between spacetimes satisfying the Julia-Nicolai condition and gravitational waves with a Killing wave vector field and, secondly, by showing the equivalence between the latter class and the one of Platonic waves.

In \cite{Julia:1994bs}, the authors focused on a class of Lorentzian manifolds which admit a null Killing vector field and which are solutions of the vacuum Einstein equations. 
In the following, we will consider spacetimes satisfying the \textit{Julia-Nicolai condition}:
$R\pl\xi,\xi\pr:=R_{\mu \nu}\xi^\mu \xi^\nu=0$, with $R$ the Ricci tensor, without the further assumption that the spacetimes considered are Einstein, as the other components of the vacuum Einstein equations play no role in the argument.
\begin{lem}[Julia-Nicolai \cite{Julia:1994bs}]
A Lorentzian manifold admitting a null Killing vector field satisfies the Julia-Nicolai condition if and only if the
null Killing vector field is hypersurface-orthogonal.
\end{lem}
%
In order to be self-contained, we review the proof presented in the section 2 of \cite{Julia:1994bs} (here in arbitrary\footnote{An alternative proof that a hypersurface-orthogonal vector field satisfies the Julia-Nicolai condition
 via the four-dimensional Raychaudhuri's equation can be found in \cite{Kramer}.} dimension) and complete some steps that were left to the reader.

\vspace{2mm}

\proof{
By contracting the commutator of two two covariant derivatives of the 1-form $\psi\equiv g\pl\xi\pr$, 
by $\xi$ and then contracting the indices
, we easily see that the Julia-Nicolai condition is equivalent to $\xi\pl\nabla^2\psi\pr:=\xi^\mu\pl g^{\rho\sigma}\nabla_\rho\nabla_\sigma\psi_\mu\pr=0$ if $\xi$ is Killing. Furthermore we have, for any Killing vector field $\xi$ with constant norm, the equivalence:
\bea\xi\pl\nabla^2\psi\pr=0\Leftrightarrow(d\psi )^2=0 \nn\eea
with $(d\psi )^2:=(d\psi )_{\mu\nu}(d\psi )^{\mu\nu}$. 
\vspace{2mm}

We now prove the following lemma:
\begin{lem}For $\xi$ a null affine geodesic vector field with dual 1-form $\psi\equiv g\pl\xi\pr$ the following equivalence holds:

$(d\psi )_{\mu\nu}(d\psi )^{\mu\nu}=0\Leftrightarrow \psi \wedge d\psi  =0$.
\end{lem}
\noindent In order to establish this lemma, we place ourselves in an adapted frame, such that the only nonvanishing component of the 1-form is $\psi _+\neq0$.\\The vector $\xi$, being affine geodesic and null, satisfies
$\xi\pl d \psi \pr=0$
 which reduces in an adapted frame to $(d \psi )_{a-}=0$ and we then have $(d \psi )^2=(d \psi )^{ij}(d \psi )_{ij}$. The condition $(d \psi )^2=0$ is then equivalent to $(d \psi )_{ij}=0$. On the other hand, the only nontrivial component of $\psi \wedge d \psi $ in this frame is $(\psi \wedge d \psi )_{+ij}=\psi _+(d \psi )_{ij}$ which also vanishes if and only if $(d \psi )_{ij}=0$, concluding the proof.
\vspace{2mm}

We therefore established the following string of equivalences:
\vspace{2mm}

\noindent $R\pl\xi,\xi\pr=0\Leftrightarrow \xi\pl\nabla^2\psi\pr=0\Leftrightarrow(d \psi )^2=0$ for a Killing vector with constant norm and $(d\psi )^2=0\Leftrightarrow \psi \wedge d\psi  =0$ which stands for a affine geodesic null vector field.
\vspace{2mm}

Remembering that the constant norm and affine geodesic conditions are satisfied by a null Killing vector field allows to write
$R\pl\xi,\xi\pr=0\Leftrightarrow\psi \wedge d\psi  =0$ for a null Killing vector field.
Using Frobenius theorem 
concludes the proof. }

We already showed in section \ref{Platonic} that Platonic waves are gravitational waves. By definition, they possess a wave Killing vector field.
Our next task concerns the equivalence of the class of Platonic waves with the class of gravitational waves with a Killing wave vector field.
\begin{prop}
A gravitational wave possesses a wave vector field that is Killing if and only if
it is a Platonic wave.
\end{prop}

\proof{
Starting from a spacetime characterized by the metric $g$ and admitting a Killing wave vector field $\xi$ (\ie $\mathcal{L}_\xi g=0$) whose dual 1-form locally reads $g\pl\xi\pr=\Om\, df$, we consider a conformally related metric $\g$ via the integrating factor $\Om$, that is $g=\Om \g$. Computing the Lie derivative
$\mathcal{L}_\xi g=\mathcal{L}_\xi\Om\,  \g+\Om\,\mathcal{L}_\xi \g$
and recalling from section \ref{Platonic screens} that the integrating factor $\Om$ of an affine geodesic wave vector field is constant along this vector field ($\mathcal{L}_\xi\Om=0$), we conclude that $\xi$ is Killing for both metrics. Furthermore, the dual 1-form associated to $\xi$ via $\g$ reads $\g\pl\xi\pr=d f$, so the vector field $\xi$ is curl-free with respect to
the metric $\g$. 
Being both Killing and curl-free, $\xi$ is parallel with respect to $\bar\nabla$ and thus, we have shown that the initial spacetime admitting a null Killing vector field is conformally related to a spacetime with respect of which this same vector becomes parallel. In other words, it is a Platonic wave.
}

From the point of view of the ambient approach,
the definition of Platonic waves as gravitational waves with a Killing wave vector field is somewhat the most natural requirement for the wavefronts to define an absolute space. Indeed, the wavefront worldvolumes are null hypersurfaces but the corresponding wavefronts or, equivalently screens, are Riemannian manifolds. The proof of this fact follows exactly the same steps as for the case of Bargmann-Eisenhart waves whose crucial ingredient was the Killing property which ensures that the metric does not depend on the choice of screen worldvolume. In other words, the Platonic waves are the most general class of gravitational waves such that their Platonic screen is canonically endowed with an Aristotelian structure\footnote{Strictly speaking, the most general class is the class of Kundt waves. More accurately, the Platonic waves are the most general waves inducing an Aristotelian structure on their space of Killing orbits. }.
\begin{prop}
The Platonic screen of a Platonic wave is a nonrelativistic spacetime, where the Aristotelian structure is induced from the ambient metric.
\end{prop}
In other words, the nonrelativistic structure of the Platonic screen is the shadow of the relativistic structure of the Platonic wave.
In a Brinkmann chart, the validity of the proposition is manifest since the absolute clock and space are respectively defined by:
$$\psi =\Omega(t,x)dt\,,\qquad d\ell^2=g_{ij}(t,x)dx^idx^j\,.$$

\subsubsection{Platonic gravitational waves as Kundt spacetimes}
\label{Plato Kundt}
We conclude this section by showing that Platonic waves belong to the Kundt class (introduced in \cite{Kundt}, see \cite{Coley:2009ut} for a detailed account), in the following sense\footnote{As for other classes of spacetimes, the terminology is a bit fuzzy in the literature because of the fact that often they are implicitly assumed to be solutions of Einstein equations (\textit{e.g.} section 27.1 of \cite{Kramer}). We adopt a geometric definition which is used for instance in \cite{Coley:2009ut}.}:
\begin{defi}
A Kundt wave is a Lorentzian manifold possessing a null geodesic, expansionless, shearless and twistless vector field.
\end{defi}
\noindent In other words, the three optical scalars of the gravitational wave must vanish.
\begin{lem}
Platonic waves are Kundt waves.
\end{lem}
This property will play an important role in the characterisation of Platonic waves (see section \ref{Further}) since the classification of Kundt waves in any dimension has recently been developed extensively \cite{Coley:2009ut}.
\proof{We already know that the null Killing vector field $\xi$ characterising a Platonic wave is hypersurface-orthogonal and geodesic. Besides, being null and Killing, the vector field $\xi$ is necessarily affine geodesic, allowing the use of the following lemma (for a proof, see \cite{Poisson} section 2.4.3):
\begin{lem}
Consider an affine geodesic vector field $\xi$, then $\xi$ is hypersurface-orthogonal if and only if its twist vanishes.
\end{lem}
Therefore the vector field $\xi$ is twistless. Furthermore, being Killing, it is also expansionless and shear-free. 
}
\textbf{Remark:} The Kundt property implies that the second fundamental form (also called extrinsic curvature) on the wavefront worldvolumes vanishes; thus, the latter are totally geodesic.
 
\noindent The general form of Kundt metrics reads \cite{Coley:2005sq}:
\bea
d\tilde{s}^2=2dt\pl du-\tilde U\pl u,t,x\pr dt+\tilde A_i\pl u,t, \vec{x}\pr dx^i\pr+\tilde g_{ij}\pl t,x\pr dx^i dx^j.  \label{Kundt}
\eea
From this canonical form of the line element, it is manifest that (i) Kundt waves are gravitational waves and (ii) Bargmann-Eisenhart waves belong to the Kundt class. The first assertion can morevover be refined as:
\bprop{A gravitational wave is a Kundt wave if and only if, in Brinkmann coordinates, the wavefront metric $\g_{ij}$ is independent of the coordinate $u$. \label{propGWKundt}}
\proof{The proof is straightforward by performing the redefinition $u\mapsto \Om\pl t,x\pr u$ in \ref{Kundt} and comparing with the line element \ref{GW}. }

However, the previously shown fact that Platonic waves belong to the Kundt class is less transparent from this point of view  
and requires additional work to make link between the canonical form of the line element for a Platonic wave \eqref{Plato} and the one for a Kundt wave \eqref{Kundt}. Starting from the Platonic line element (\ref{Plato}) and performing the redefinition $u'=\Om \,u$ puts the Platonic metric in the Kundt form (\ref{Kundt}) with $\tilde U \pl u',t,x\pr=\Om\pl t, x\pr \bU\pl t, x\pr+u'\p_t\pl \ln \Om\pr$ and $\tilde A_i\pl u',t,x\pr=\Om\bA_i\pl t,x\pr-u'\p_i \pl \ln \Om\pr$. The potential and Coriolis form acquire a linear dependence in $u'$ and then Platonic waves are seen to belong to the more restrictive class of \textit{degenerate} Kundt spacetimes \cite{Coley:2009ut} for which the potential and Coriolis form of (\ref{Kundt}) take the specific form 
\footnote{A more geometric definition of degenerate Kundt spacetimes states that a degenerate Kundt wave has to satisfy the following two conditions: i) it must be a Kundt wave with respect to a null vector $\ell$ and ii) the Riemann tensor and all its covariant derivatives must be of type II (or more special) in the kinematic (\textit{i.e.} aligned with $\ell$) frame, see section \ref{Further} for terminology. }: 
\bea
\tilde U\pl u,t,x\pr&=&u^2\tilde U^{(2)}\pl t,x\pr+u\, \tilde U^{(1)}\pl t,x\pr+\tilde U^{(0)}\pl t,x\pr\label{degK}\\
\tilde A_i\pl u,t,x\pr&=&u\, \tilde A_i^{(1)}\pl t,x\pr+\tilde A_i^{(0)}\pl t,x\pr\nn.
 \eea
By comparison with the transformed $\tilde U$ and $\tilde A_i$, we see that for a Platonic wave brought in the canonical degenerate Kundt form \eqref{Kundt}-\eqref{degK}, we have $\tilde U^{(2)}=0$, $\tilde U^{(1)}=\p_t\pl \ln \Om\pr$ and $\tilde U^{(0)}=\Om\bU$ as well as $\tilde A_i^{(1)}=-\p_i \pl \ln \Om\pr$ and $\tilde A_i^{(0)}=\Om \bA_i$.
\begin{prop}
Platonic waves are degenerate Kundt waves.
\end{prop}
The coordinate transformations 
\bea
u&=&u'\pl\frac{\p t}{\p t'}\pr^{-1}+f\pl t',\vec x'\pr\nn\\
t&=&t\pl t'\pr\nn\\
x^{i}&=&x^{i}\pl t',\vec x'\pr\nn
 \eea
together with the redefinitions
\bea
U^{'(2)}\pl  t',\vec x'\pr&=&U^{(2)}\nn\\
U^{'(1)}\pl t',\vec x'\pr&=&U^{(1)}\frac{\p t}{\p t'}-A_i^{(1)}\frac{\p x^i}{\p t'}+2f U^{(2)}\frac{\p t}{\p t'}\nn\\
U'^{(0)}&=&\frac{\p t}{\p t'}\Bigg[\crl U^{(0)}+fU^{(1)}+f^2U^{(2)}\crr\frac{\p t}{\p t'}+\pl\frac{\p t}{\p t'}\pr^{-2}\frac{\p^2 t}{\p t^{'2}}-\frac{\p f}{\p t'}\nn\\
&&- \crl A_i^{(0)}+fA_i^{(1)}\crr \frac{\p x^i}{\p t'}\Bigg]-\half g_{ij}\frac{\p x^i}{\p t'}\frac{\p x^j}{\p t'}\nn\\
A_i^{'(1)}
&=&A_j^{(1)}\frac{\p x^j}{\p x^{i'}}\nn\\
A_i^{'(0)}&=&\frac{\p t}{\p t'}\crl  \frac{\p f}{\p x^{i'}}+\pl A_j^{(0)}+fA_j^{(1)}\pr \frac{\p x^j}{\p x^{i'}}\crr+g_{kl}\frac{\p x^k}{\p t'}\frac{\p x^l}{\p x^{i'}}\nn\\
 g_{ij}'&=&g_{kl}\frac{\p x^k}{\p x^{i'}}\frac{\p x^l}{\p x^{j'}}\nn.
  \eea
preserve the canonical form of the line element \eqref{Kundt}-\eqref{degK} for a degenerate Kundt wave. Remarkably, these transformations also preserve the subclass of Platonic waves written in the canonical form of degenerate Kundt waves in the sense that  $\tilde U^{'(2)}=0$, $\tilde U^{'(1)}=\p^{'}_t\pl \ln \Om\pr$ as well as $\tilde A_i^{'(1)}=-\p^{'}_i \pl \ln \Om\pr$. This fact will be useful in the future proof of proposition \ref{CSIPlato}.

Finally, we summarize the hierarchy of properties that have been discussed in the following chain of inclusion:
\bea
&\mbox{Gravitational waves}&\nn\\
&\bigcup&\nn\\
&\mbox{Kundt waves}&\nn\\
&\bigcup&\nn\\
&\mbox{Degenerate Kundt waves}&\nn\\
&\bigcup&\nn\\
&\mbox{Platonic waves}&\nn\\
&\bigcup&\nn\\
&\mbox{Bargmann-Eisenhart waves}&\nn\\
&\bigcup&\nn\\
&\mbox{pp-waves}&\nn
\eea
\subsection{Miscellaneous Platonic waves}
\label{Miscellaneous}

As an illustration, we now briefly review various types of proper Platonic waves (\textit{i.e.} which do not belong to the Bargmann-Eisenhart class).
\vspace{2mm}

\noindent\textbf{Anti de Sitter spacetime:} the most symmetric example of a proper Platonic wave. The existence of a null Killing vector field is manifest in the Poincar\'e coordinates 
\bea\label{AdSlinel}
ds^2=\,\frac1{z^2}[2\, dudt+\,dz^2\,+\,d\vec y^2].
\eea
As one can see, the wavefronts are hyperbolic spaces of dimension $d$ as is manifest from their line element: $d\ell^2=\frac1{z^2}[dz^2\,+\,d\vec y^2]$. In other words, anti de Sitter (AdS) spacetime is an example of a Lobachevsky-plane-fronted wave.
\vspace{2mm}

\noindent\textbf{AdS-gyraton \cite{Frolov:2005ww}:} Lobachevsky-plane-fronted wave conformally equivalent to a pp-wave whose line element writes
\bea
ds^2&=&\,\frac1{z^2}\crl2\,dt\pl du-\bU(t,z,\vec y)dt+\bA_i\pl t,z,\vec y \pr\pr \,+\,dz^2\,+\,d\vec y^2\crr.
\eea
All curvature scalar invariants of AdS-gyratons are constant and identical to the ones of AdS.

\noindent\textbf{Siklos spacetime \cite{Siklos}:} Coriolis-free AdS-gyratons of line element
\bea
ds^2=\,\frac1{z^2}\crl2\,dt\pl du-\bU(t,z,\vec y)dt\pr \,+\,dz^2\,+\,d\vec y^2\crr. 
\eea
This definition is related to one of the equivalent characterization of the class of ``Lobachevsky-plane gravitational wave'' by Siklos himself in $D=4$ dimensions \cite{Siklos}. They were later reinterpreted as ``AdS pp-waves'' in \cite{Podolsky:1997ik}.
Siklos waves are Kerr-Schild spacetimes \textit{i.e.} can be written as $g_{\mu \nu}=\mathfrak g_{\mu \nu}-2{\mathfrak U}\xi_\mu\xi_\nu$ with $\mathfrak g$ the AdS metric.  In Brinkmann coordinates the background metric reads \eqref{AdSlinel} while the Kerr-Schild potential writes ${\mathfrak U}=z^2\,  \bU$ and $\xi=\frac{\p}{\p u}$ is the null Killing vector field.
Siklos spacetimes are Einstein if and only if the scalar potential $\bU$ has vanishing Laplace-Beltrami operator on AdS space, \textit{i.e.} $\frac{1}{\sqrt {-\mathfrak g}}\p_\mu\pl\sqrt {-\mathfrak g}\mathfrak g^{\mu\nu}\p_\nu \bU\pr=z^2\pl\p^2_z\bU+\p_i\p^i\bU\pr+\pl2-D\pr z\p_z\bU=0$. 
Einstein Siklos waves are furthermore weakly universal \cite{Coley:2008th}, as will be discussed in section \ref{Further}.
\vspace{2mm}

\noindent\textbf{Kaigorodov solution \cite{Kaigorodov}:} Siklos spacetime with potential that only depends on the coordinate $z$ 
(in the Brinkmann-Poincar\'e coordinates) in the following way: $\bU(z)\propto z^n$ (with $D=n+1$ the dimension of spacetime).
Without loss of generality, its line element is thus
\bea
ds^2=\,\frac1{z^2}\crl2\,dt\pl du\pm z^n\,dt\pr \,+\,dz^2\,+\,d\vec y^2\crr. 
\eea
Kaigodorov solutions belong to the class of Einstein Siklos spacetimes. In other words, they are vacuum solutions in the presence of a negative cosmological constant.

\vspace{2mm}

\noindent\textbf{Schr\"odinger spacetime (Sch$_Z$):} Siklos spacetime where, in the Brinkmann-Poincar\'e coordinates, $\bU(z)\propto z^{2(1-Z)}$ where $Z\geqslant1$ is called the \textit{dynamical exponent} because of the nonrelativistic scale transformation $t\mapsto \lambda^Z t$, $\vec x\mapsto \lambda\, \vec x$, with $\vec x: =(z,\vec y)$ and  $u\mapsto \lambda^{2-Z} u$, which preserves the line element
\bea
ds^2=\,\frac1{z^2}\crl2\,dt\pl du+z^{2(1-Z)}dt\pr \,+\,dz^2\,+\,d\vec y^2\crr. 
\eea
Anti de Sitter spacetime corresponds to $Z=1$: Sch$_1$ = AdS which is the homogeneous manifold for the isometry group $O(n,2)$ acting on its conformal boundary as conformal transformations. From the point of view of symmetries, the dynamical exponent $Z=2$ is also of high interest: Sch$_2$ is a homogeneous manifold (see
\cite{Duval:2012qr,Hartong:2012sw} for detailed global and coordinate-independent descriptions) with the Schr\"odinger group $Sch(d)$ as the isometry group that acts on the conformal boundary as Schr\"odinger transformations (this was the property that motivated their introduction in \cite{Son:2008ye}).
Contrary to Kaigorodov solutions, the Schr\"odinger spacetimes Sch$_Z$ for $Z\neq 1$ are not solutions of Einstein equations, even in the presence of a cosmological constant. However, they are solutions of richer theories with exotic matter (such as Proca fields \cite{Son:2008ye}) or some supergravity theories (see \eg references in \cite{Hartong:2012sw}).

\begin{figure}[ht]
\centering
   \includegraphics[width=0.8\textwidth]{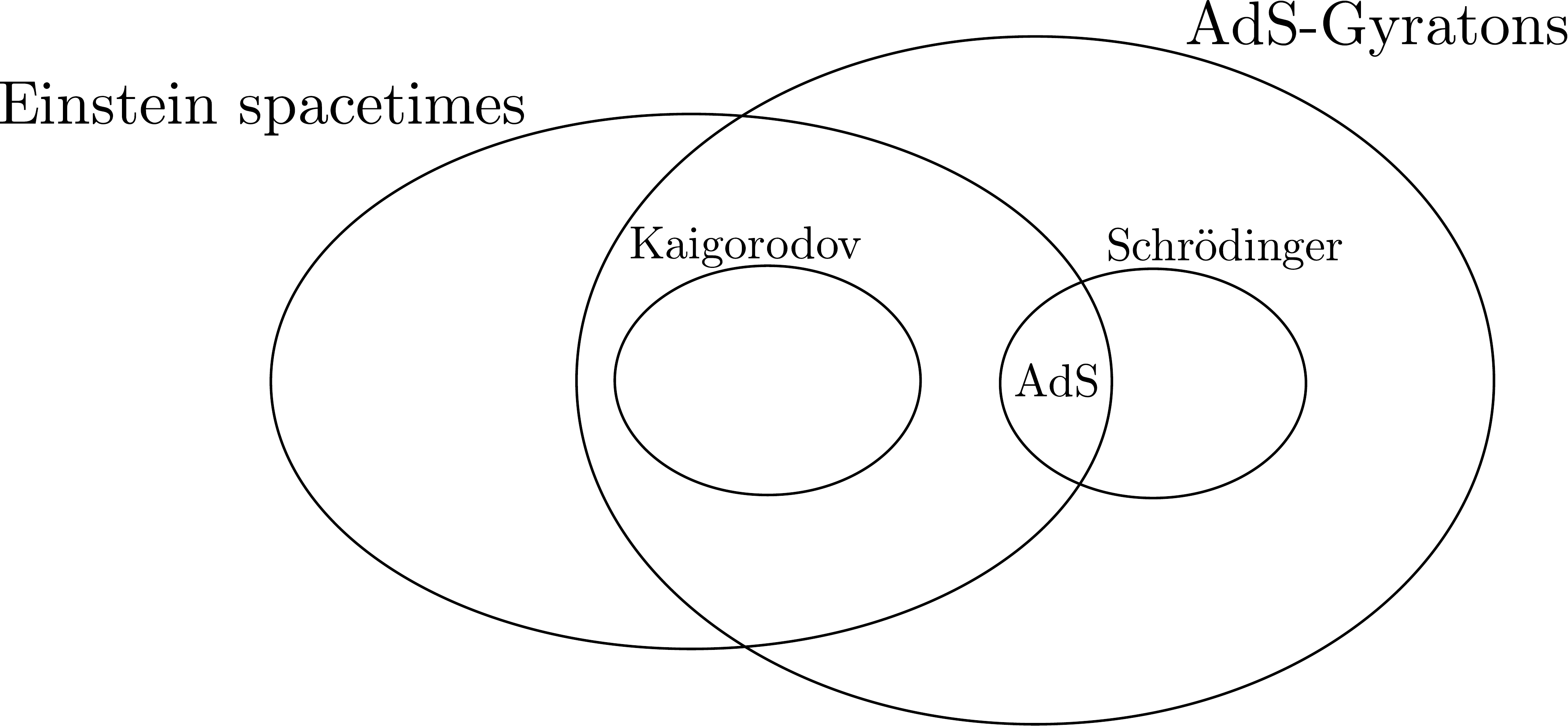}
  \caption{AdS-gyratons}\label{fig8}
\end{figure}

We summarize in figure \ref{fig8} the main class of Platonic examples whose physical interest is well established by the considerable literature dwelled upon.

\vspace{2mm}
\noindent\textbf{Platonic plane waves:} 
Spacetimes whose line element reads:
\bea
ds^2=\frac{1}{\vec x^2}\crl 2dt\pl du-\bU\pl t, \vec x\pr dt+\bA_i\pl t, \vec x\pr dx^i\pr+d\vec x^2\crr\,.
\eea
This name has been chosen because they are indeed plane-fronted Platonic waves in $D=4$ (and their wavefronts are cylinders $\mathbb R\times \mathbb S^{d-1}$ in higher dimensions), as can be seen in the spherical coordinates with radial coordinate $r=|\vec x|$.
The $D=4$ dimensional Platonic plane waves form the only class of nonhomogeneous plane-fronted proper Platonic spacetimes with constant scalar curvature invariants, as will be explained in section \ref{Further}. However, they seem of little physical interest since none of them are Einstein manifolds.

\section{Geometric properties of Platonic gravitational waves}
\label{Geometric properties Plato}

\subsection{Global properties: 
completeness and causality}
\label{Global}

Since global issues are investigated in the present section, one should be more specific about the global structure of the spacetimes which will be considered. For the sake of simplicity, we will restrict our analysis to Platonic waves with:
\begin{enumerate}
	\item[(i)] topology ${\mathbb R}^2\times\Sigma$, where ${\mathbb R}^2$ corresponds to the domain of $(u,t)$ in the Brinkmann coordinates,
	\item[(ii)] conformal factor $\Omega$ and components $g_{\alpha\beta}$ of the spacetime metric that are regular functions of $t$ and $x^i$,	
	\item[(iii)] geodesically complete wavefronts $\Sigma$ endowed with the metric $g_{ij}$,
	\item[(iv)] conformally related Bargmann-Eisenhart waves such that their wavefronts $\Sigma$ are endowed with a time-independent metric ${\bar g}_{ij}$ and are geodesically complete.
\end{enumerate}

Physically, an important property of spacetimes is the absence of singularities, in the sense of geodesic completeness.
Effectively, the geodesics of Platonic waves are described as trajectories for a dynamical system \eqref{lagrang} defined in 
terms of the components of the metric ${\bar g}_{ij}$, the vector potential $\bA_i$ and the effective potential $\bar V$. 
Due to the above simplifying assumptions, the only way for a geodesic to be incomplete in this restricted class of Platonic waves is that the corresponding dynamical trajectory goes to spatial infinity in a finite 
time. 

Heuristically, one might expect that the radial behavior of the effective potential at spatial infinity controls the motion of observers at large distances, so that the behaviors of the conformal factor and scalar potential would control the geodesic completeness of Platonic waves.
Indeed, these ideas can be converted into a theorem, which is a perfect example of the utility of the ambient approach in the study of gravitational waves. Its proof is essentially a byproduct of Eisenhart-Lichnerowicz theorem, \textit{i.e} the geodesic completeness of Platonic waves
follows from the completeness of the corresponding dynamical trajectories. 
For Bargmann-Eisenhart waves, this is an equivalence \cite{Candela:2002rr}.
The distinction arises for proper Platonic waves (\ie $\Omega \neq$const) because of the fact that finite time intervals $\Delta t=t_1-t_0$ along a dynamical trajectory always correspond to
finite affine-parameter intervals 
\bea\Delta \tau=\tau_1-\tau_0=\frac1{m}\int_{t_0}^{t_1}{\Omega\big(t,x(t)\big)\,dt}\eea along an ambient geodesic, since by assumption (ii) the conformal factor $\Omega$ is finite for any value of $t$ and $x^i$. However, the converse is not necessarily true because if $\Omega$ tends to zero when $|t|\to\infty$, then $\Delta \tau$ may be finite even for infinite $|\Delta t|$.

In order to state our result, some definitions should be introduced. 
Let us denote by
\bea
\Vert x\Vert=\int\limits_{x_0}^x \sqrt{\bg_{ij}(x')\, dx^{'i} dx^{'j}} 
\eea the geodesic distance from the ``origin'' (chosen to be any given point) $ x_0$ on $\Sigma$.
``Spatial infinity'' corresponds to the limit $\Vert x\Vert\to \infty$.
\begin{defi}[Candela, Romero, S\'anchez \cite{Candela:2012}]
A function $f(t,x)$ on ${\mathbb R}\times\Sigma$ grows at most quadratically along finite times if for each $T>0$ there exist some positive constants $A_T$ and $B_T$ such that $$f(x,t)\,\leqslant\,A_T\Vert x\Vert^2+B_T\qquad \forall (t,x)\in [-T,T]\times\Sigma\,.$$
The function is said to grow subquadratically along finite times if the inequality is strict.
\end{defi}

A corollary\footnote{Their corollary was not stated with the same degree of generality as formulated here, though the authors of \cite{Candela:2002rr,Candela:2012} must be aware of this stronger formulation since it follows in a straightforward way from their many results.}
of the works \cite{Candela:2002rr,Candela:2012}
is the following fact:
\begin{prop}[Candela, Flores, Romero, S\'anchez \cite{Candela:2002rr,Candela:2012}]
Bargmann-Eisenhart waves obeying to conditions (i)-(iii)
and with potential $\bU(x,t)$
decreasing [\textit{i.e.} $-\bU(x,t)$ growing] at most quadratically  at spatial infinity along finite times are geodesically complete.
\end{prop}

Therefore, by merely adapting the powerful results of \cite{Candela:2012} (in particular theorem 2) on the completeness of dynamical trajectories, one can show:
\begin{prop}\label{Pgeodcomp}
Platonic waves obeying to conditions (i)-(iv) with:
\begin{itemize}
 \item conformal factor $\Omega(t,x)$,
 \item minus the scalar potential $-\bU(x,t)$, 
 \item absolute value of the time derivative of the conformal factor $|\partial_t \Omega(t,x)|$,
 \item absolute value of the time derivative of the scalar potential $|\partial_t \bar U(t,x)|$,
\end{itemize}
that grow at most quadratically at spatial infinity along finite times,
are geodesically complete.
\end{prop}
One should stress that the above bounds on the growths are with respect to the geodesic distance on $\Sigma$ defined by the spatial metric $\g_{ij}$ (so \textit{not} by the wavefront metric $ g_{ij}=\Omega \g_{ij}$).
\proof{Ambient geodesics with $m=0$ are effectively described as geodesics of the wavefronts $\Sigma$ with respect to the metric $g_{ij}$. They are ensured to be complete by hypothesis (iii). 

Ambient geodesics with $m\neq 0$ are effectively described as dynamical trajectories with respect to the action principle \eqref{action}. Theorem 2 of \cite{Candela:2012} applies because of hypotheses (i)-(iv) and ensures that they are complete if minus the effective potential $-\bar V$ and the absolute value of its time derivative $|\partial_t \bar V|$ grow at most quadratically along finite times. Indeed, the effective potential $\bar V=\bar U +\half\,\frac{M^2}{m^2}\,\Omega$, defined by \eqref{UbarVbar}, decreases at most quadratically at finite times for all values of $M^2\in\mathbb R$ because of the four hypotheses on the growing behavior.
 Similarly, $|\partial_t\bar V|\leqslant |\partial_t\bar U| +\half\,|\frac{M^2}{m^2}|\,|\partial_t\Omega|$ grows at most quadratically at finite times.}

 \noindent\textbf{Application:}
Schr\"{o}dinger spacetimes Sch$_Z$ with $Z\geqslant 2$ 
are expected to be geodesically complete gravitational waves, as follows from the above proposition. This remains obscure in the local Poincar\'e-like coordinates but becomes more manifest in the global ``trapping'' coordinates
\bea
ds^2=\frac{1}{z^2}\crl2dt\pl du-\half\Big( \cos^{2\pl Z-2\pr}\pl t\pr z^{2\pl1-Z\pr}+ z^2+\vec y^2\Big) dt+dz^2+d\vec y^2\pr\crr
\eea
introduced in \cite{Blau:2009gd} for this purpose. The Schr\"{o}dinger spacetimes with $Z=2$ were proved to be geodesically complete in \cite{Blau:2009gd} but the case $Z>2$ was left open.
The domain $0<z<\infty$ fulfills the assumptions (i)-(iii) for $Z>2$ (this condition ensures the regularity of the scalar potential). The conformal factor and scalar potential satisfy the hypotheses of the proposition \ref{Pgeodcomp} for $Z>5/2$. Indeed, for all $Z>1$, $\Om$ and $\p_t\Om$ go to zero when $z$ goes to $\infty$ and $-\bU<0$. Moreover, $|\p_t\bU|=\tl\pl Z-2\pr\cos^{2Z-5}\pl t\pr\sin\pl t\pr z^{2\pl1-Z\pr}\tr$ grows at most quadratically in $z$ for $Z>5/2$. Strictly speaking, the assumption (iv) is not satisfied because $\g_{ij}=\delta_{ij}$ is the flat metric and the half-space $0<z<\infty$ is not geodesically complete since straight lines may cross the boundary $z=0$. Nevertheless, this subtlety should not be a problem in regard of the geodesic completeness taking into account the known fact from \cite{Blau:2009gd} that, for $Z\geqslant 2$, timelike and lightlike geodesics cannot reach $z=0$ for a finite value of the affine parameter. Still, this fact prevents us from a full rigorous proof of the geodesic completeness for $Z>2$.\footnote{The upper bound $Z>5/2$ can be optimized till $Z>2$ by adapting the corollary 3 of \cite{Candela:2012}.}

\vspace{2mm}

Another important global property of spacetimes is their causal structure.
By definition, a Platonic wave is conformally related to a Bargmann-Eisenhart wave; thus, both share locally the same causal structure. Therefore, without loss of generality one may restrict the study of causal properties of the Platonic waves to the one of Bargmann-Eisenhart waves, being careful about the domain of definition of the conformal map.
Platonic waves are causal spacetimes \cite{Flores:2002fx} but not more in general.\footnote{We refer to the sections 4.1 of the review \cite{GarciaParrado:2005yz} for a useful reminder of the hierarchy of causality conditions in general relativity. 
} For instance, a celebrated result of Penrose is his proof \cite{Penrose:1965rx} that exact plane waves are strongly causal but not globally hyperbolic (nor causally simple). As a byproduct of the ambient approach, the property of causal simplicity of Bargmann-Eisenhart waves was shown to be equivalent (modulo technical assumptions) to the existence of maximizers for the proper time between causally related events \cite{Minguzzi:2006gq}.

As geodesic completeness, the causal structure of Platonic waves is governed by the behavior of the potential at spatial infinity.
Indeed, the following theorem was shown \cite{Flores:2002fx} for Bargmann-Eisenhart waves ${\mathbb R}^2\times\Sigma$ which are Coriolis-free
and with time-independent geodesically-complete wavefronts $\Sigma$: if the potential decreases, at spatial infinity,
with respect to the Riemannian distance on the wavefront
(I) at most quadratically, then it is strongly causal, or (II) subquadratically, then it is globally hyperbolic.
There is a wide class of relevant gravitational waves which satisfy the assumption (I) but not (II) and which are geodesically complete and strongly causal but not globally hyperbolic.
Exact plane wave solutions and anti de Sitter spacetimes are the perfect example of such Platonic waves.

As one can see, the faster the potential decreases, the weaker is the causal structure of the Platonic wave. In fact, another result of \cite{Flores:2002fx} for these same generic classes of spacetimes is that: if the potential is nonpositive and decreases superquadratically (\textit{i.e.} faster than $-\Vert x\Vert^2$) at spatial infinity, then it is not distinguishing (which is the weakest condition coming after mere causality).
In any case, an important lesson to draw is that Platonic waves should be such that their scalar potential is bounded from below or at most decreases slowly at spatial infinity in order to have standard causality properties and no singularity.

Finally, because of the importance of black objects in contemporary general relativity, another important global issue
is the existence of an event horizon.
Partial answers are that Coriolis-free pp-waves cannot possess a horizon while some examples of Platonic waves do possess one
\cite{Hubeny:2003ug}. However, such black waves are generated by somewhat exotic matter and it has been shown that
a large class of Platonic waves with a regular horizon cannot be solutions of Einstein equations in vacuum or with null matter \cite{Liu:2003cta}.

\subsection{Curvature scalar invariants: classification}
\label{Further}

Curvature scalar invariants (\textit{i.e.} scalars built as polynomials formed from the Riemann tensor and its covariant derivatives) constitute a powerful tool in the equivalence problem, that is the task to determine if two given metrics are locally isomorphic or not. As such, Riemannian manifolds are entirely determined by their curvature scalar invariants  \cite{Hervik:2010rg} and one is then able to tell if two Riemannian manifolds are isomorphic by systematically comparing their respective curvature scalar invariants. For Lorentzian spacetimes though, this theorem does not hold and there exists a nontrivial class of spacetimes which are not uniquely characterized by their invariants so that more elaborate procedures such as the Cartan-Karlhede algorithm are needed in order to solve the equivalence problem. In four dimensions, this special class of spacetimes is identified with the one of degenerate-Kundt metrics \cite{Coley:2009eb}, introduced in section \ref{Plato Kundt}, so that nonequivalent degenerate Kundt metrics can share identical invariants. Although it stays true that degenerate-Kundt spacetimes are not determined by their scalar invariants in higher dimensions \cite{Coley:2010ze}, it remains to be proved that they are the only higher dimensional spacetimes enjoying this property. We established earlier that Platonic waves are degenerate-Kundt; therefore, we formulate the following:
\begin{prop}
Platonic waves are not determined by their scalar curvature invariants.
\end{prop}
The very existence of a class of spacetimes not being characterized by their invariants opens the possibility of Lorentzian manifolds having vanishing curvature scalar invariants (called VSI spacetimes in the following) without necessarily being flat. As is obvious from the previously stated theorem, the only Riemannian VSI manifolds are flat. By definition, curved Lorentzian VSI manifolds are not determined by their scalar curvature invariants 
and, furthermore, it can be shown that they belong to the degenerate-Kundt class in any dimension \cite{Coley:2006fr}. The authors of \cite{Coley:2004hu} showed that in arbitrary dimension a spacetime is VSI if and only if it belongs to the Kundt class, \textit{i.e.} admits a geodesic nonexpanding, shear-free and twist-free null vector field $\xi$, and the Riemann tensor is of type III (or more special) relative to $\xi$. The second condition
involves the notion of the boost order of a tensor, which we define, following the terminology introduced in \cite{Coley:2004jv} (see \cite{Reall:2011ys} for a pedagogical review), as the difference between the number of ``$+$'' and ``$-$'' in the components of a covariant tensor (concretely, all down indices) written in an adapted frame. The condition prescribing that the Riemann tensor of a VSI spacetime must be of type III relative to $\xi$ is equivalent to have a Riemann tensor with strictly negative boost order when computed in the adapted frame.

Concretely, 
the condition that the boost order of the Riemann tensor
is strictly negative amounts to the set of equations below:
\bea
\begin{array}{|c|c|} \hline\mbox{Boost order} & \mbox{Riemann component}\nn\\
\hline
2&R_{+i|+j}=0\\
1&R_{+-|+i}=R_{+i|jk}=0\\
0&R_{+-|+-}=R_{+-|ij}=R_{+i|-j}=R_{ij|kl}=0\\
\hline
\end{array}
\eea
We now focus on the VSI spacetimes among the Platonic waves and prove the following lemma:
\begin{lem}
A Bargmann-Eisenhart wave is VSI if and only if it is a pp-wave.
\end{lem}
\proof{
Platonic waves belong to the Kundt class so
the only remaining condition to satisfy is that the boost order of the Riemann tensor
is negative.

The existence of a congruence of parallel rays implies that
$R_{ab|cd}\xi^d=0$ and thus we have, in the kinematic frame, $R_{ab|c+}=0$ since $\xi$ is nowhere vanishing. Therefore the first six conditions are
automatically satisfied for a Bargmann-Eisenhart wave. The last condition is equivalent to be plane fronted.
}
The extension of this result to Platonic waves is rendered quite simple by the useful result of \cite{Coley:2007yx} stating that if a VSI spacetime admits a null (or timelike) Killing vector field $\xi$, then $\xi$ is necessarily parallel. Therefore the class of VSI Platonic waves reduces to the one of VSI Bargmann-Eisenhart spacetimes and we have the following proposition:
\begin{prop}
A Platonic wave is VSI if and only if it is a pp-wave.
\end{prop}
\noindent One way to heuristically interpret this result is to consider that the VSI property of a Platonic wave descends to the wavefront, which being Riemannian, must necessarily be flat.
\vspace{2mm}

We now consider the natural extension of the VSI class that is spacetimes possessing \textit{constant} curvature scalar invariants (CSI). For Riemannian manifolds, the class of CSI metrics reduces to (locally) homogeneous manifolds \cite{Prufer1996}. The Lorentzian case is again richer as, in four dimensions, the CSI class is composed of all (locally) homogeneous manifolds as well as a subset of the degenerate Kundt spacetimes dubbed \textit{degenerate-CSI$_K$} metrics \cite{Coley:2009tx}.
Degenerate-CSI$_K$ are Kundt spacetimes for which there exists a frame such that all curvature tensors (that is the Riemann tensor and all its covariant derivatives) have vanishing positive boost weight components and constant boost weight zero components.
In higher dimensions, the situation is less clear than in the VSI case as it is not yet known if the class of (locally) homogeneous spacetimes together with the class of degenerate-CSI$_K$ spacetimes exhaust the CSI class when $D>4$. For this reason, we will focus in the sequel on the $D=4$ case.

\begin{figure}[ht]
\centering
\includegraphics[width=0.7\textwidth]{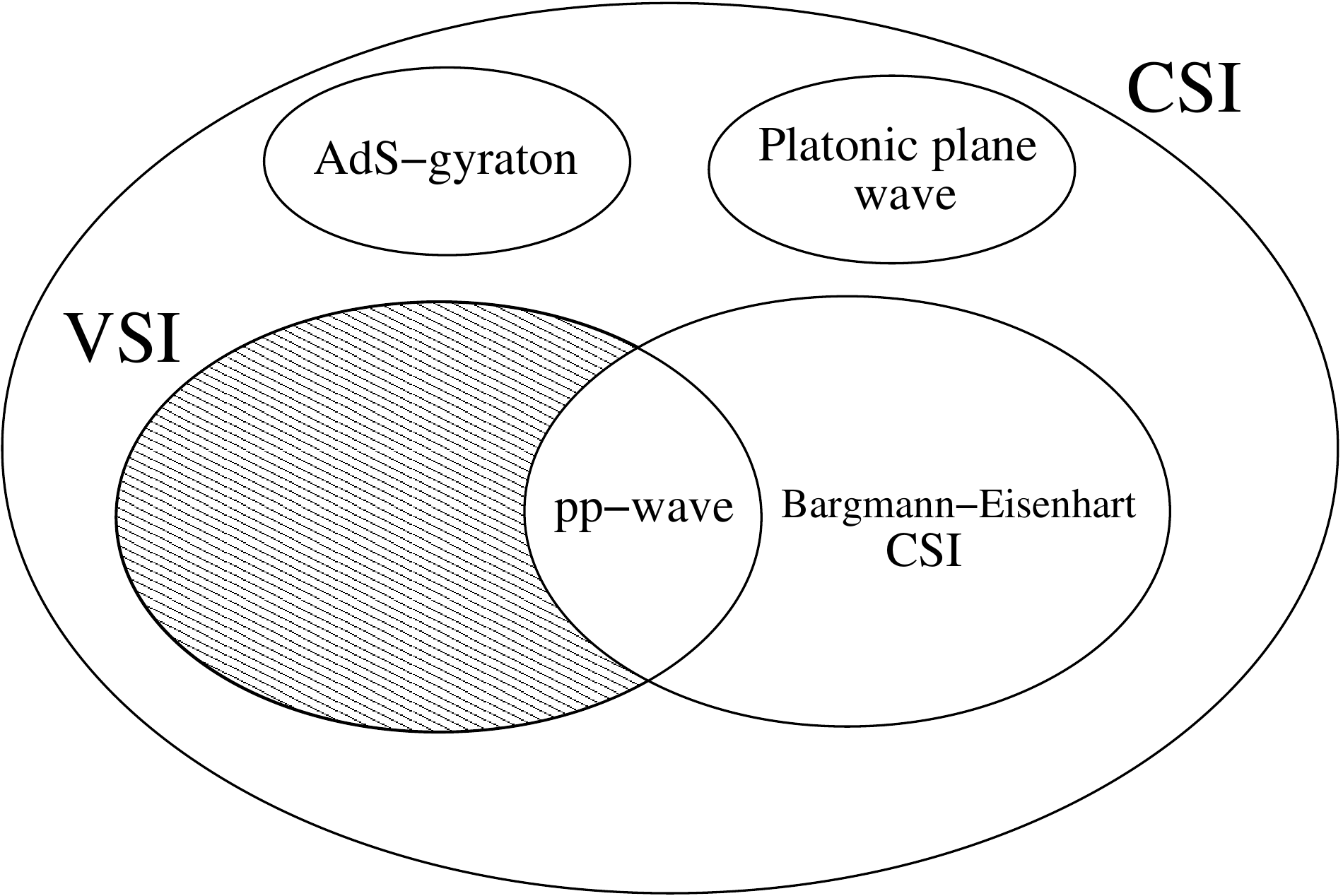}
\caption{\label{etiquette} Four-dimensional Platonic CSI spacetimes: Note that the set VSI$\smallsetminus$pp-wave is empty. }
\end{figure}

\noindent We again start with the Bargmann-Eisenhart case. Actually, this question has already been addressed in \cite{Mcnutt:2009zz} (where Bargmann-Eisenhart spacetimes are denoted CCNV) and the following proposition has been established:
\begin{prop}[McNutt, Coley, Pelavas \cite{Mcnutt:2009zz}]
A four-dimensional Bargmann-Eisenhart wave is CSI if and only if its wavefront is locally homogeneous.
\end{prop}
Again, we note that the CSI property seems to befall to the wavefront. There are three types of 2-dimensional locally homogeneous Riemannian spaces, respectively locally isometric to: the sphere $\mathbb S^2$, the Euclidean plane $\mathbb E^2$ and the hyperbolic plane $\mathbb H^2$. The general expression of a four-dimensional CSI Bargmann-Eisenhart spacetime, in Brinkmann coordinates then reads:
\bea
ds^2=2\,dt\pl du-\bU(t,\vec x)dt+\bA_i(t,\vec x)\, dx^i\pr \,+d\ell^2
\eea
where the wavefront line element takes the form
$d\ell^2=dx^2+\frac{1}{\lambda^2}\sin^2\pl\lambda x\pr dy^2$ where ${\mathbb S}^2$: $\lambda^2>0$, $\mathbb E^2$: $\lambda^2=0$ and $\mathbb H^2$: $\lambda^2<0$.
Obviously, the Euclidean case corresponds to a pp-wave and the spacetime is then VSI.
In order to address the Platonic case, we will rely on the classification of four dimensional degenerate-CSI$_K$ metrics proposed in \cite{Coley:2009tx} and prove the following proposition:
\begin{prop}\label{CSIPlato}
A four-dimensional Platonic wave is CSI if and only if it belongs to one of the following classes:
\begin{itemize}
\item locally homogeneous
\item CSI Bargmann-Eisenhart
\item AdS-gyraton
\item Platonic plane wave.
\end{itemize}
\end{prop}
\proof{As stated earlier, four-dimensional CSI spacetimes consist of all locally homogeneous or degenerate-CSI$_K$ spacetimes\footnote{Note that these two classes intersect, see \textit{e.g.} footnote $14$ in \cite{Ortaggio:2012jd}.}. We now focus on Platonic waves belonging to the degenerate-CSI$_K$ class and make use of the classification of four-dimensional degenerate-CSI$_K$ displayed in \cite{Coley:2009tx}. More technically, the authors of \cite{Coley:2009tx} wrote, for each class of locally homogeneous wavefront (\textit{i.e.} ${\mathbb S}^2$, $\mathbb E^2$ and $\mathbb H^2$) the two-dimensional 1-forms $\tilde A^{(1)}_i$ allowing the construction of a degenerate-CSI$_K$ spacetime. Our task is then to require that the obtained line element matches the form of Platonic waves seen as degenerate-Kundt metrics (see section \ref{Plato Kundt}) for some function $\Omega\pl t,x\pr$. This requirement is quite drastic as, besides the CSI Bargmann-Eisenhart, only two classes of proper Platonic waves remain, namely AdS-gyratons and Platonic plane waves. }
We note that no nonhomogeneous spherical wavefront proper 
Platonic waves are CSI. The figure \ref{etiquette} provides a summary of the Platonic CSI spacetimes.
\vspace{2mm}

The study of CSI spacetimes is partly motivated by the physically relevant notion of ``universality'' which designates the property enjoyed by spacetimes which are vacuum solutions of any theory of quantum gravity (in the sense of effective field theory, \eg the string theory low-energy effective action). A more precise definition \cite{Coley:2008th} distinguishes between weakly (and strongly) universal spacetimes to designate spacetimes for which any conserved symmetric tensor of rank two constructed from the metric, the Riemann tensor and its covariant derivatives is a constant multiple of the metric (vanishes). The link between the universality and CSI properties has been highlighted in \cite{Coley:2011dn}, where it was shown that any universal four-dimensional spacetime must be CSI. However, there is still no crisp result allowing us to discriminate which CSI spacetimes are universal. A conjectured candidate for a subset of universal CSI spacetimes are the so-called CSI$_\Lambda$ spacetimes \cite{McNutt:2012bj} whose invariants constructed from the traceless Ricci tensor, Weyl tensor and their covariant derivatives vanish. The work \cite{McNutt:2012bj} displays a classification of four-dimensional CSI$_\Lambda$ spacetimes which relies on the one proposed in \cite{Coley:2009tx}. Then, by similar arguments as the one used in the proof of proposition \ref{CSIPlato}, we establish the following fact:
\begin{prop}
CSI$_\Lambda$ Platonic waves are either pp-waves or AdS-gyratons.
\end{prop}
Indeed, this class contains the two classes of universal Platonic waves already known in the literature: Coriolis-free pp-waves have been shown to be strongly universal in \cite{Horowitz:1989bv} while Siklos waves are known to be weakly universal \cite{Coley:2008th}.

\section{Conclusion}

In this work, 
we investigated how nonrelativistic physics can be embedded inside relativistic gravitational waves.

We started by reviewing the work of Eisenhart and Lichnerowicz 
on null dimensional reduction of geodesics to nonrelativistic dynamical trajectories, first in the Lagrangian framework and then in the more suitable Hamiltonian formalism, where the deep mechanism behind the Eisenhart lift was shown to appear in a more transparent way.
Moreover, the quantum analogue of the ambient approach was addressed: the Schr\"odinger equation on curved space was obtained from the Klein-Gordon equation for a free scalar field on curved spacetime via null dimensional reduction.

Then we focused on the ambient approach to gravity by first showing how relativistic gravitational waves could induce a nonrelativistic (\ie Aristotelian) structure on their Platonic screen. Gravitational waves have indeed been shown to allow a natural definition of a (locally synchronizable) absolute clock, although they generally lack the structure necessary to induce an absolute space. Bargmann-Eisenhart waves were 
introduced 
and the arguments of \cite{Duval:1984cj} were reproduced in order to show that waves possessing a parallel null vector field do induce a well-defined Aristotelian structure on their Platonic screen. 
These results were extended to the larger class of Platonic waves, seen as gravitational waves admitting a Killing wave vector field, considered in \cite{Julia:1994bs}.
They were shown to constitute the most general gravitational waves inducing an Aristotelian structure on any screen worldvolume.
Meanwhile, we provided a new geometric definition for Platonic waves, as conformal Bargmann-Eisenhart waves admitting the same null Killing vector field, of which we made substantial use in various proofs.

As first applications, the results of \cite{Candela:2002rr,Candela:2012} concerning the geodesic completeness of Bargmann-Eisenhart waves were extended to Platonic waves and, as a corollary, evidence was provided that Schr\"odinger manifolds with dynamical exponent greater than two are geodesically complete.
As a second application, the classification of Platonic waves with constant curvature scalar invariants was addressed.
Although the extension of the class of Bargmann-Eisenhart waves to the one of Platonic waves does not allow the inclusion of new  spacetimes with vanishing scalar invariants, it does enlarge the class of spacetimes with constant curvature scalar invariants
allowing the Eisenhart lift.
Namely, it includes the degenerate-CSI$_K$ classes formed by AdS-gyratons and Platonic plane waves. We also considered the more restricted class of CSI$_\Lambda$ spacetimes and established that the only Platonic CSI$_\Lambda$ are pp-waves and AdS-gyratons. 
The link with the class of spacetimes which are vacuum solutions of any gravity theory was also briefly discussed. 

It would be interesting to push further the ambient approach to gravity by generalizing to Platonic waves the null dimensional reduction of Einstein equations, performed in \cite{Duval:1984cj} for Bargmann-Eisenhart waves, by making use of the definition we introduced and to compare with the analysis of \cite{Julia:1994bs}. A formulation in terms of Cartan's connection might also shed some light on the origin of Newtonian connections, as in \cite{Kunzle}. This would also provide a basis for a generalization of the ambient approach to Vasiliev higher-spin gravity and a possible check of the holographic duality proposed in \cite{Bekaert:2011cu}.

\section*{Acknowledgements}

We are very grateful to C.~Duval and P.~Horv\'athy for useful exchanges and discussions on nonrelativistic structures.
We also wish to thank M.~Grigoriev and M.~Henneaux for discussions on constrained Hamiltonians, C.~Barrabes and P.~Hogan on gravitational waves. We further acknowledge D.~McNutt, M.~S\'anchez and V.~Pravda for correspondence and helpful comments on their respective works.



\end{document}